\documentclass[citeautoscript,floatfix,aps,prb,twocolumn,superscriptaddress]{revtex4-1}

\usepackage{latexsym}
\usepackage{graphicx}
\usepackage{hyperref}
\usepackage[version=4]{mhchem}
\usepackage{amsmath,amssymb,amsfonts}
\usepackage{color}
\usepackage{appendix}
\usepackage[dvipsnames]{xcolor}
\usepackage[T1]{fontenc}
\usepackage[english]{babel}
\usepackage{url}
\usepackage{bm}
\usepackage[utf8]{inputenc}
\usepackage{xr}
\usepackage{changes}

\newcommand{\be}{\begin{equation}}
	\newcommand{\ee}{\end{equation}}

\definecolor{forestgreen}{RGB}{34,139,34}

\usepackage{todonotes}
\definecolor{coral}{RGB}{200,86,60}

\begin{document}

	\title{Electronic Structure of Chromium Trihalides beyond Density Functional Theory}
	\author{Swagata Acharya}
	\email{swagata.acharya@ru.nl}
	\affiliation{Institute for Molecules and Materials, Radboud University, {NL-}6525 AJ Nijmegen, The Netherlands}	
	\author{Dimitar Pashov}
	\affiliation{ King's College London, Theory and Simulation of Condensed Matter,
		The Strand, WC2R 2LS London, UK}
	\author{Brian Cunningham}
	\affiliation{Centre for Theoretical Atomic, Molecular and Optical Physics, Queen’s University Belfast, Belfast BT71NN, Northern Ireland, United Kingdom}
	\author{Alexander N. Rudenko}
	\affiliation{Institute for Molecules and Materials, Radboud University, {NL-}6525 AJ Nijmegen, The Netherlands}
	\author{Malte R\"{o}sner}	
	\affiliation{Institute for Molecules and Materials, Radboud University, {NL-}6525 AJ Nijmegen, The Netherlands}
	\author{Myrta Gr\"{u}ning}	
	\affiliation{Atomistic Simulation Centre, Queen’s University Belfast, Belfast BT71NN, Northern Ireland, United Kingdom}	
	\author{Mark van Schilfgaarde}
	\affiliation{ King's College London, Theory and Simulation of Condensed Matter,
		The Strand, WC2R 2LS London, UK}
	\affiliation{National Renewable Energy Laboratory, Golden, CO 80401, USA}	
    \author{Mikhail I. Katsnelson}
	\affiliation{Institute for Molecules and Materials, Radboud University, {NL-}6525 AJ Nijmegen, The Netherlands}

	\begin{abstract}
		
		We explore the electronic band structure of free standing monolayers of chromium trihalides, CrX\textsubscript{3}{, X= Cl,
			Br, I}, within an advanced \emph{ab-initio} theoretical approach based in the use of Green's function functionals. 
			We compare the local density approximation with the
		quasi-particle self-consistent \emph{GW} approximation (QS\emph{GW}) and its self-consistent extension (QS$G\widehat{W}$) by solving the particle-hole
		ladder Bethe-Salpeter equations to improve the effective interaction \emph{W}.  We show that at all levels of theory, the valence band consistently changes
		shape in the sequence Cl{\textrightarrow}Br{\textrightarrow}I, and the valence band maximum shifts from the M point to
		the $\Gamma$ point.  However, the details of the transition, the one-particle bandgap, and the eigenfunctions change
		considerably going up the ladder to higher levels of theory.  The eigenfunctions become more directional, and at the M point there
		is a strong anisotropy in the effective mass. Also the dynamic and momentum dependent self energy shows that QS$G\widehat{W}$ adds to the
		localization of the systems in comparison to the QS\emph{GW} thereby leading to a narrower band and reduced amount of halogens in the valence band manifold.
		
	\end{abstract}
	
	\maketitle
	
	With the discovery of ferromagnetic order in \ce{CrI3}, the family of chromium trihalides \ce{CrX3}, X=Cl, Br, I, has
	emerged as a new class of magnetic 2D crystals. Ferromagnetism (FM) in a monolayer \ce{CrI3} was
	first reported in 2017~\cite{huang,klein}, which was followed by observation of FM in
	CrBr$_{3}$~\cite{zhang,kim}, CrCl$_{3}$~\cite{cai} and many other compounds~\cite{li,fei,deng,gong,gibert}.  FM is
	intrinsic to these system, which distinguishes them from traditional 2D $sp$-electron magnets where magnetism is induced by proximity
	to a FM substrate.  Long range-order is typically suppressed in two-dimensional magnets~\cite{mermin}, but it can be
	stabilized by magneto-crystalline anisotropy, which opens an energy gap in the magnon spectra and therefore protects
	the FM in two dimensions~\cite{IKK,soriano_review}. Due to their layered structure and their weak inter-layer van-der-Waals interactions these systems are loosely coupled to their substrates, which provides greater
	flexibility in functionalizing them and controlling their properties, e.g. by varying the layer-number or by applying a gate voltage.  This offers new possibilities to make spintronic devices with high accuracy and
	efficiency~\cite{jiang,shan,jia,kolekar,mak,song,song1,ubrig,yang,klein}.

	\begin{figure}[h]
		\begin{center}
			\includegraphics[width=0.54\columnwidth]{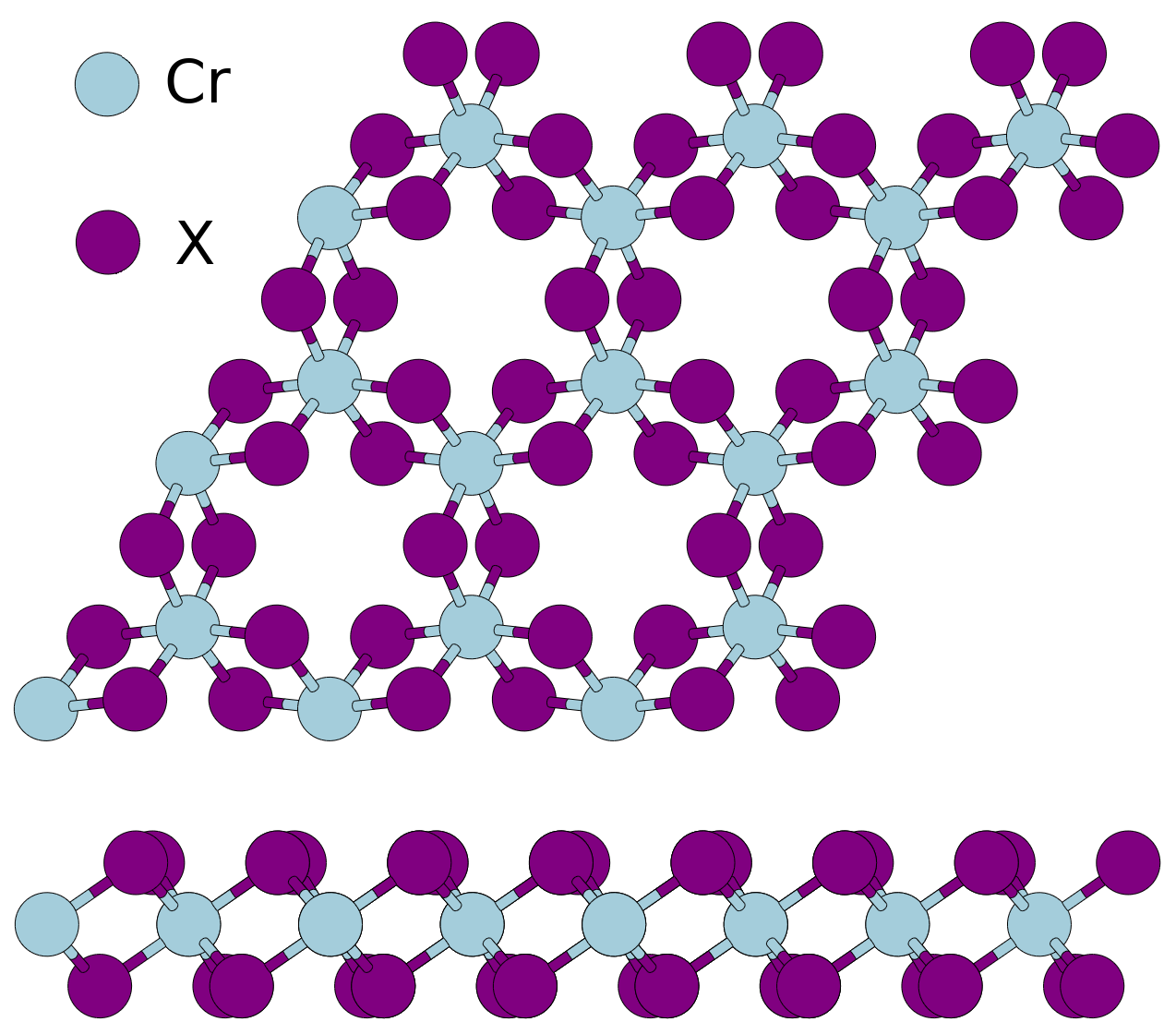}
			\includegraphics[width=0.43\columnwidth]{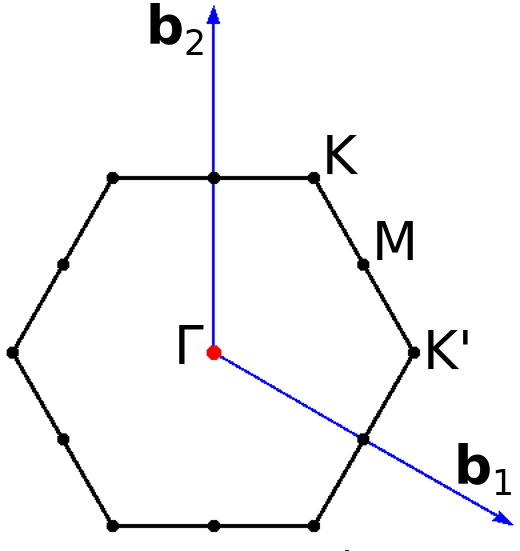}
			\caption{Left: Ball-and-stick model of the crystal structure of monolayer chromium trihalides Cr$X_3$ ($X$=Cl, Br, I). Right: Brillouin zone of the corresponding hexagonal lattice with the high-symmetry points indicated. ${\bf b}_1$ and ${\bf b}_2$ denote reciprocal lattice vectors. 
}
			\label{fig:struc}
		\end{center}
	\end{figure}

	CrX$_{3}$ is a two-dimensional FM insulator with FM originating from the Cr-X-Cr
	superexchange interaction~\cite{soriano_review,kashin2020,kvashnin2020,soriano_model}.  Six Cr$^{3+}$ ions form a honeycomb structure with D$_{3d}$ point group symmetry and each Cr is surrounded by six X in an octahedral geometry (see Fig.~\ref{fig:struc}). The edge-sharing geometry leads to first neighbor Cr atoms sharing a pair of ligands. This enables pathways for Cr-X-Cr super-exchange.  In this crystal field geometry, the Cr \emph{d} splits into a t$_{2g}$ triplet and an e$_{g}$
	doublet.  Cr$^{3+}$ has a valence of three electrons, which fill the t$_{2g}$ majority-spin band
	according to Hund’s first rule, leaving all other \emph{d} bands empty.  The ionic model leads to the magnetic moment
	on each Cr$^{3+}$ ion of $\sim$3$\mu_{B}$ which is confirmed by ab-initio calculations.  
	All three \ce{CrX3} compounds have FM order down to the monolayer with Curie temperatures T$_{I}$ = 45 K~\cite{huang}, T$_{Br}$ = 34 K~\cite{zhang} and T$_{Cl}$ = 17 K~\cite{cai} and the magnetization easy axis is normal to the plane for \ce{CrI3} and \ce{CrBr3} while it is in plane for \ce{CrCl3}.

	Recent Density Functional Theory (DFT) calculations~\cite{molina,khomskii,soriano} confirm the qualitative understandings derived from the ionic model. 
	However, at quantitative level details start to differ from the fully ionic picture; one important such factor is the degree of hybridization of the t$_{2g}$ levels with the \emph{p} bands of the ligands. This degree of hybridization depends on the ligand, its atomic weight and the number of core levels, which turns out to be a crucially important factor in determining the detailed
	electronic band structure.  This is the main focus of the present paper which we carefully analyze on different levels of theory beyond conventional density functional theory. Our self-consistent $GW$ (QS\emph{GW}) and $BSE$ (QS$G\widehat{W}$) implementations are independent of the starting point and, hence, allow us to study the roles of self-consistent charge densities and self-energies in determining the key features of the electronic structures at different levels of the theory. \\

	\noindent\emph{Molecular Picture}
	
	Within the local-density approximation (LDA), we find the spin-polarized bandgaps of the three systems to be, 1.51 eV, 1.30
	eV and 1.20 eV for X=(Cl, Br, I) respectively, in line with prior work~\cite{molina}.  The qualitative trend
	is easily understood in terms of the splitting between Cr $d$ and X $p$ atomic levels.  In the simplest two-level
	tight-binding description, the conduction and valence levels are given by
	$(\varepsilon_d{+}\varepsilon_p)/2{\pm}\sqrt{((\varepsilon_d{-}\varepsilon_p)/2)^2 + v^2}$, where $\varepsilon_d$ and
	$\varepsilon_p$ are respectively the Cr $t_{2g}$ \emph{d} and X \emph{p} atomic levels and $v$ the hybridization matrix
	element.  This results in a gap $E_{g}=\varepsilon_d-\varepsilon_p+2v^2/(\varepsilon_d-\varepsilon_p)$ to the lowest order in
	$v/(\varepsilon_d-\varepsilon_p)$.  Among all three halides Iodine has the most core levels resulting in its $p$ levels having the most
		nodes which thus sense the attractive nucleus most weakly.
	It has the shallowest $\varepsilon_p$ of the three halogens, while
	Cl has the deepest.  Thus, the qualitative trend in the bandgap is simply understood as following from the hallide $\varepsilon_p$ energies relative to the Cr $\varepsilon_d$.  As a slight elaboration on this
	picture that includes magnetism, we can distinguish between the majority ($t_{2g}^\uparrow$) and minority
	($t_{2g}^\downarrow$) Cr $d$ levels.  For the bandgap, the picture just sketched corresponds to the
	($\varepsilon_{d}^\downarrow-\epsilon_p$) bond.  A similar picture applies to the $\varepsilon_{d}^\uparrow-\epsilon_p$
	bond, but in this channel both bond and antibond are occupied, and moreover $\varepsilon^\uparrow_d-\epsilon_p$ need not
	be large in comparison to $v$.  Indeed the Cr $t_{2g}^{\uparrow}$ and X $p$ levels may overlap.\\
	
	\begin{figure}[t]
		\begin{center}
			\includegraphics[width=1.02\columnwidth]{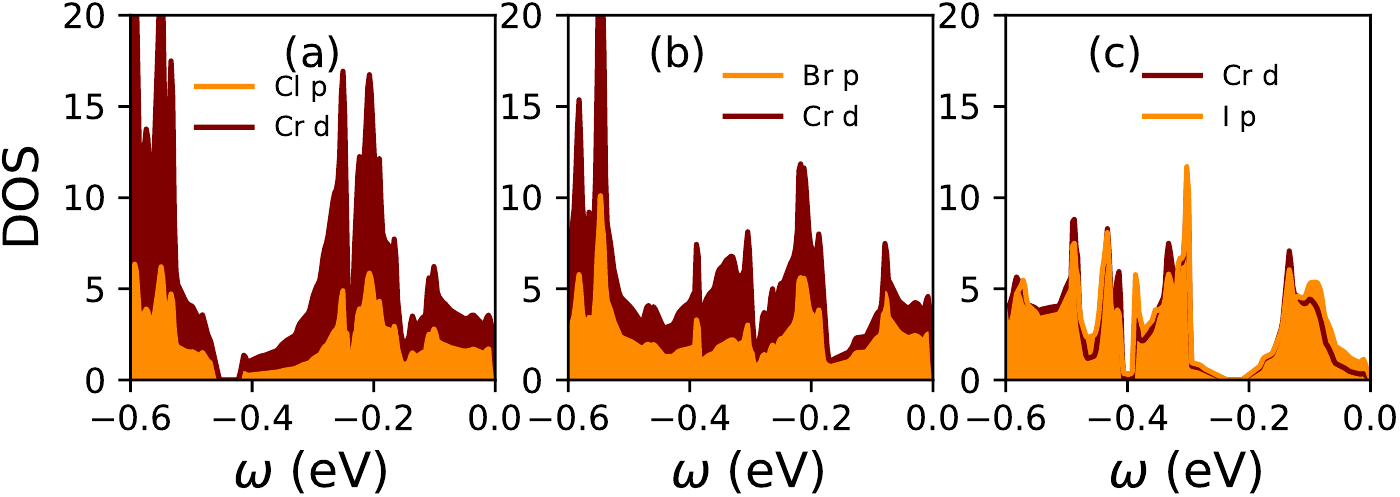}
			\caption{Partial Density of states within the LDA, projected onto the Cr-\emph{d} and X-\emph{p}
				states for (a) X=Cl, (b) Br and (c) I.}
			\label{fig:ldados}
		\end{center}
	\end{figure}
	
	\begin{table}[h]
		\footnotesize
		\begin{tabular}{c|@{\hskip 3pt}ccc@{\hskip 3pt}|@{\hskip 3pt}ccc}
			\hline
			theory & \multicolumn{3}{c}{bandgap (eV)} & \multicolumn{3}{c}{spectral weight} \cr
			& \ce{CrCl3} & \ce{CrBr3}  & \ce{CrI3} & \ce{CrCl3} & \ce{CrBr3} & \ce{CrI3} \cr
			\hline
			DFT    & 1.51  & 1.30   & 1.06   & 24\% & 31\% & 50\%  \cr
			QSGW   & 6.87  & 5.73   & 3.25   & 45\% & 69\% & 84\%  \cr
			QS$G\widehat{W}$    & 5.55  & 4.65   & 2.9    & 27\% & 37\% & 71\%  \cr
		\end{tabular}
		\caption{One particle electronic band gap at different levels of theory (with spin-orbit coupling). The gap increases from LDA to QS\emph{GW} level. When
			when ladder diagrams are added two-particle interactions via a BSE, $W{\rightarrow}{\widehat{W}}$ and screening is increased.
			This reduces the QS\emph{GW} bandgap.  Right columns show fraction of spectral weight that the Halogen
			contributes to the total DOS within an energy window of occupied states ($-$0.6,0)\,eV, relative to the valence band maximum.}
		\label{tab:gaps}
	\end{table}
	
	\noindent\emph{Energy band picture}
	
	
	The molecular picture qualitatively explains the trends in the
	bandgap and the admixture of X $p$ in the highest valence states in the sequence Cl{\textrightarrow}Br{\textrightarrow}I. However, in the 2D crystal, the molecular levels broaden into bands which can alter the
	trends in both the bandgap and the merging of X $p$ with Cr $t_{2g}^{\uparrow}$ in the valence bands. The corresponding orbital resolved density of states are shown in
	Fig.~\ref{fig:ldados}.  The X \emph{p} level becomes more shallow, and the highest lying valence band acquires
	increasing anion character as can be seen both in Table~\ref{tab:gaps} and in Fig.~\ref{fig:ldados}.  Spin-orbit
	coupling only slightly modifies the electronic structure for \ce{CrCl3} and \ce{CrBr3}, while for \ce{CrI3} the bandgap
	reduces by 150 meV to 1.06 eV in the LDA.
	
\begin{figure}
		\begin{center}
			\includegraphics[width=0.34\columnwidth, angle=-0]{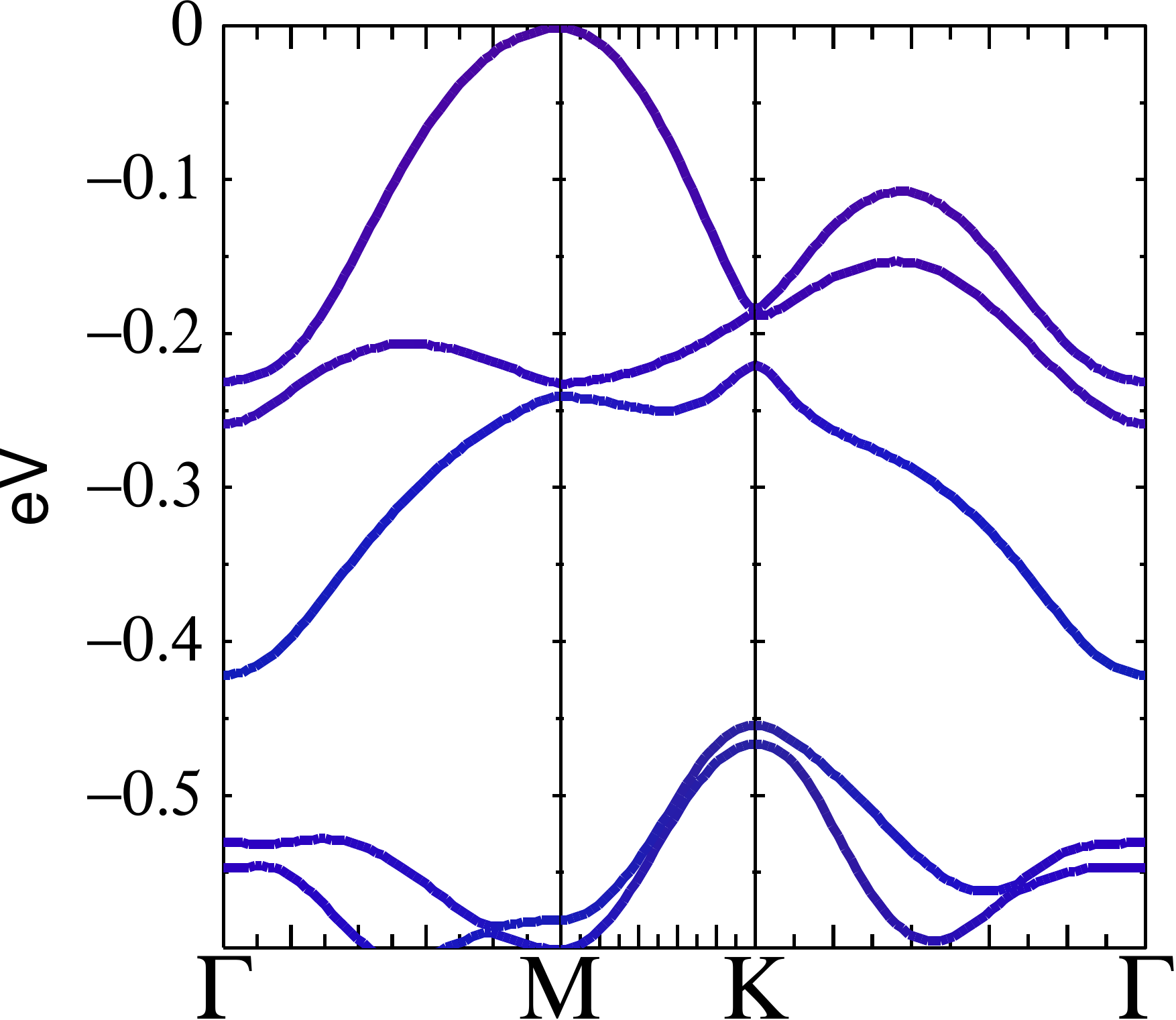}
			\includegraphics[width=0.318\columnwidth, angle=-0]{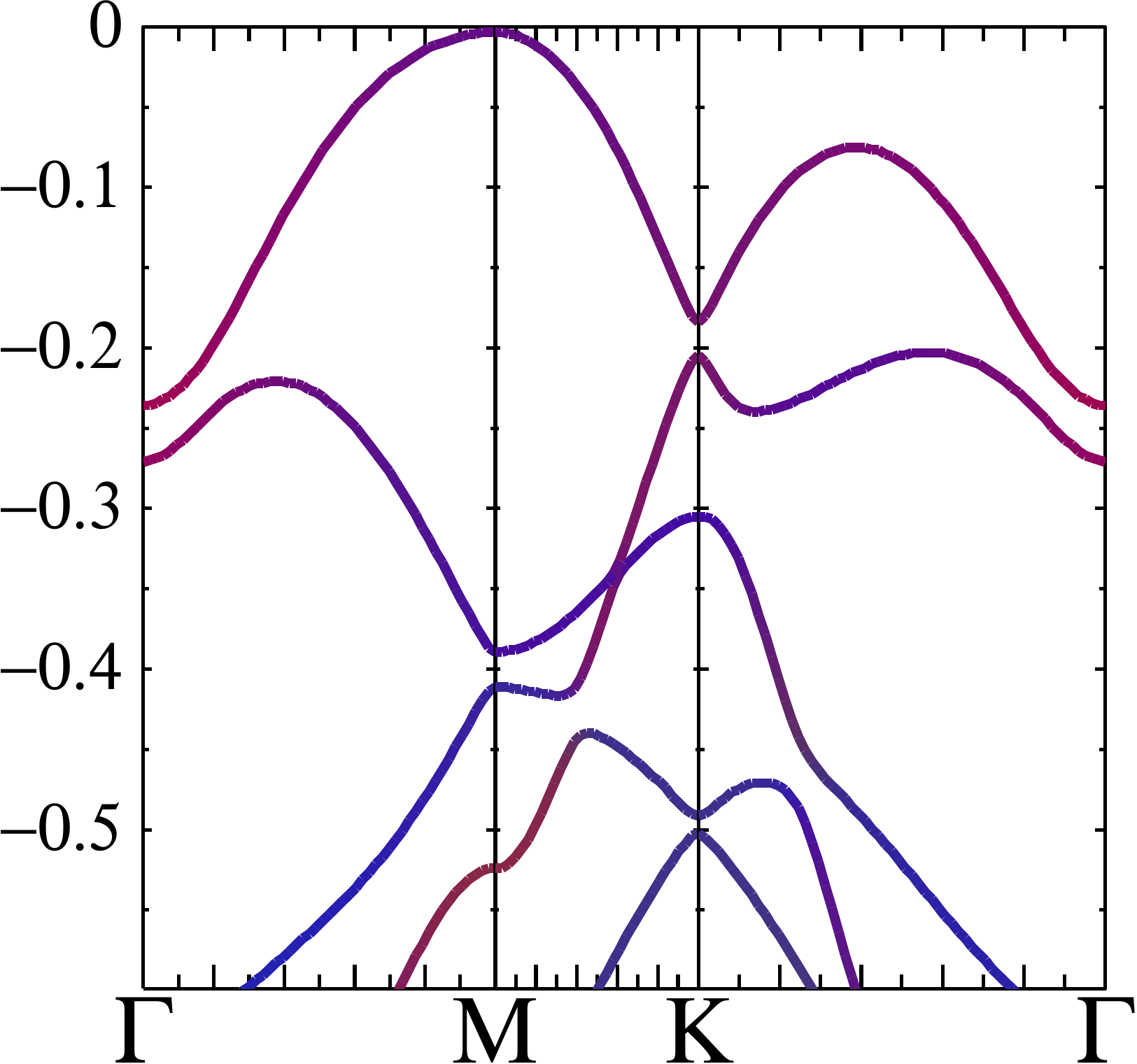}
			\includegraphics[width=0.318\columnwidth, angle=-0]{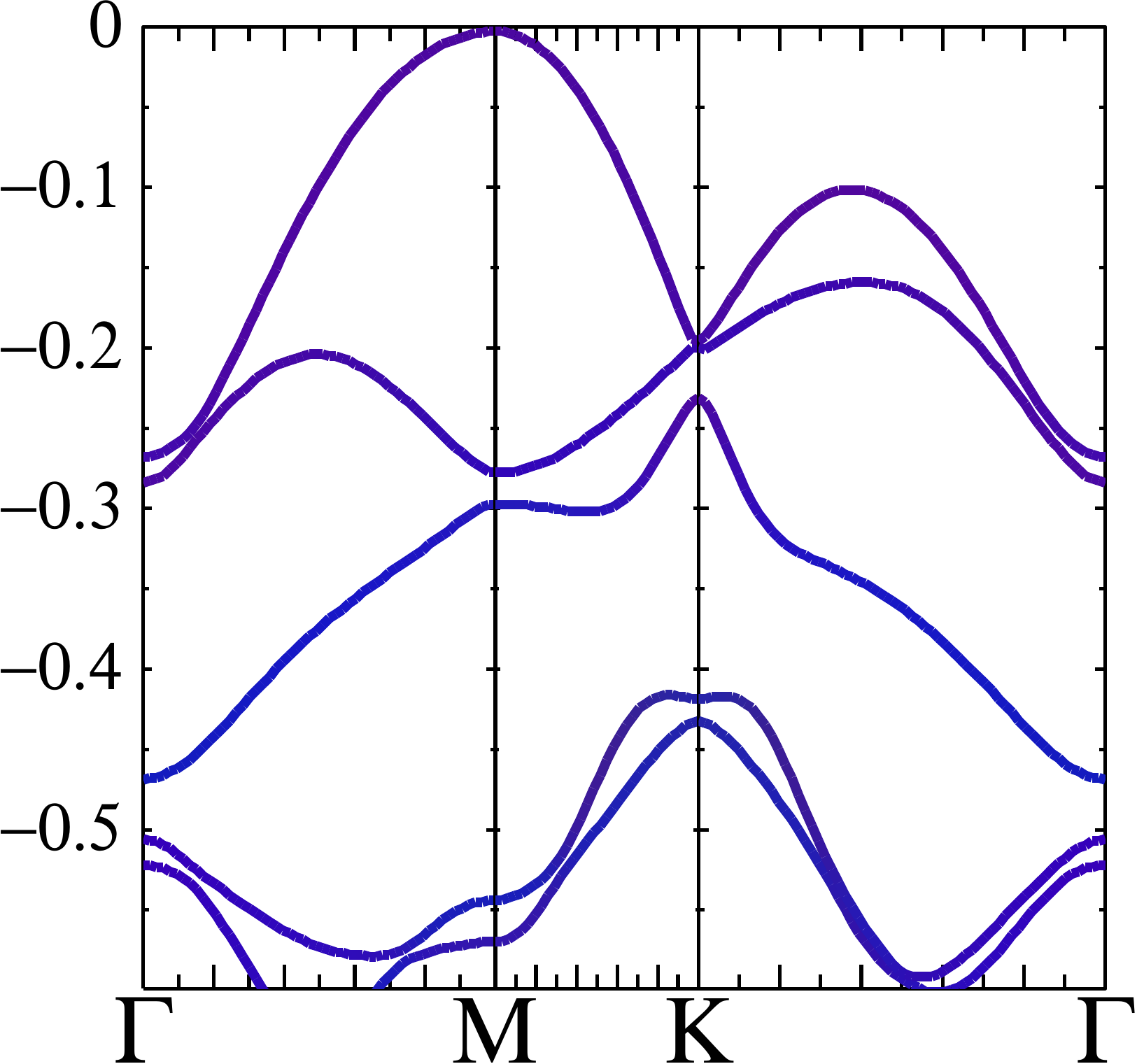}\\
			\includegraphics[width=1.05\columnwidth]{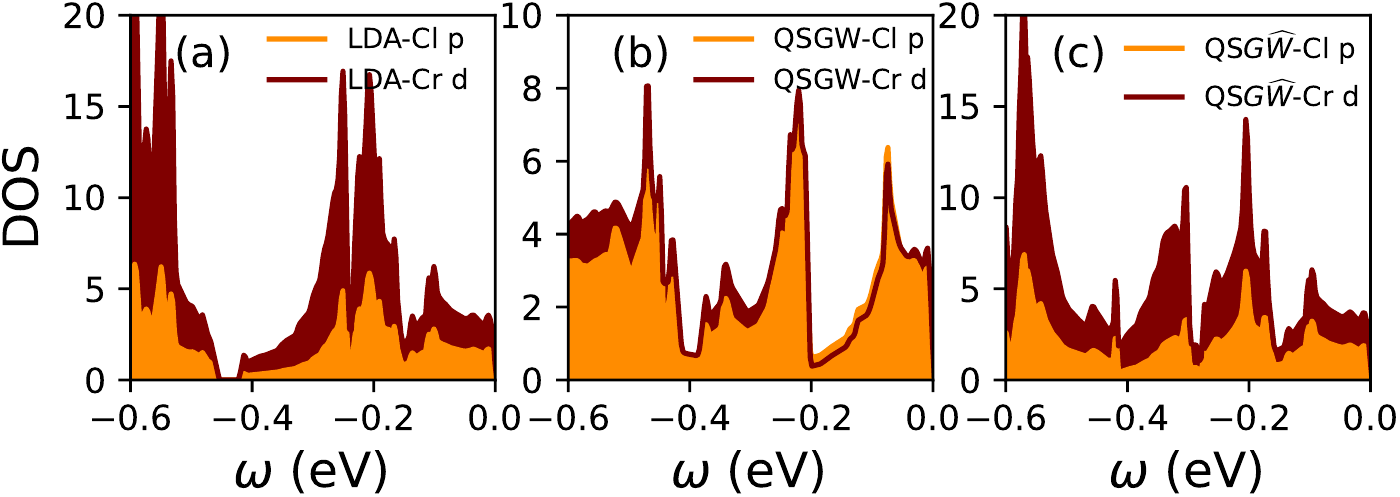}
			\caption{\ce{CrCl3} : From left to right: LDA, QSGW and QS$G\widehat{W}$ valence band structures (with spin-orbit coupling) are shown
				respectively. The colors correspond to Cl-$p_{x}+p_{y}$ (red), Cl-p$_{z}$ (green), Cr-d (blue). In the
				bottom panels are shown the density of states projected onto the Cr-d and Cl-p states at different
				levels of the theory.}
			\label{fig:cl}
		\end{center}
	\end{figure}

\begin{figure}
		\begin{center}
			\includegraphics[width=0.34\columnwidth, angle=-0]{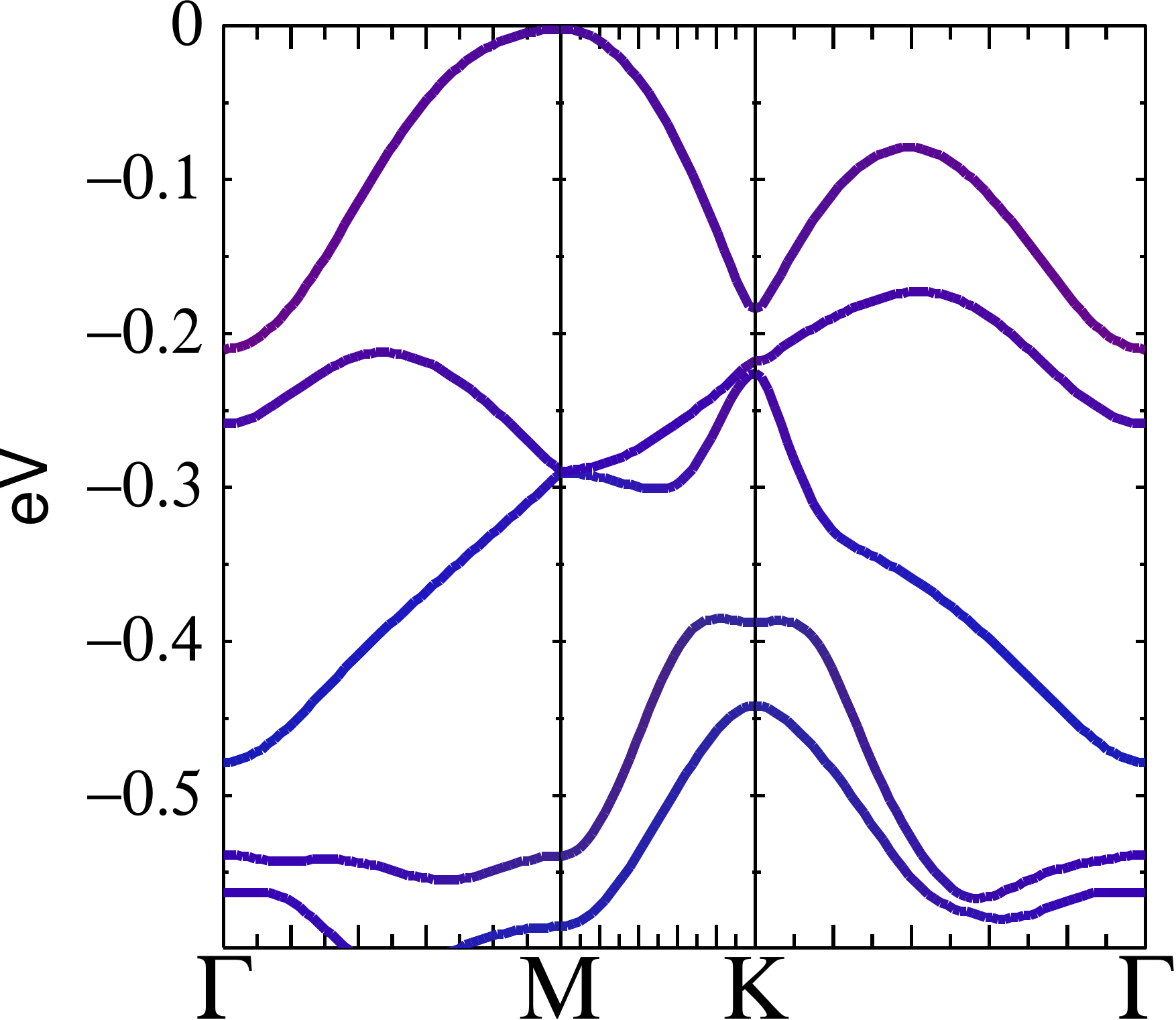}
			\includegraphics[width=0.318\columnwidth, angle=-0]{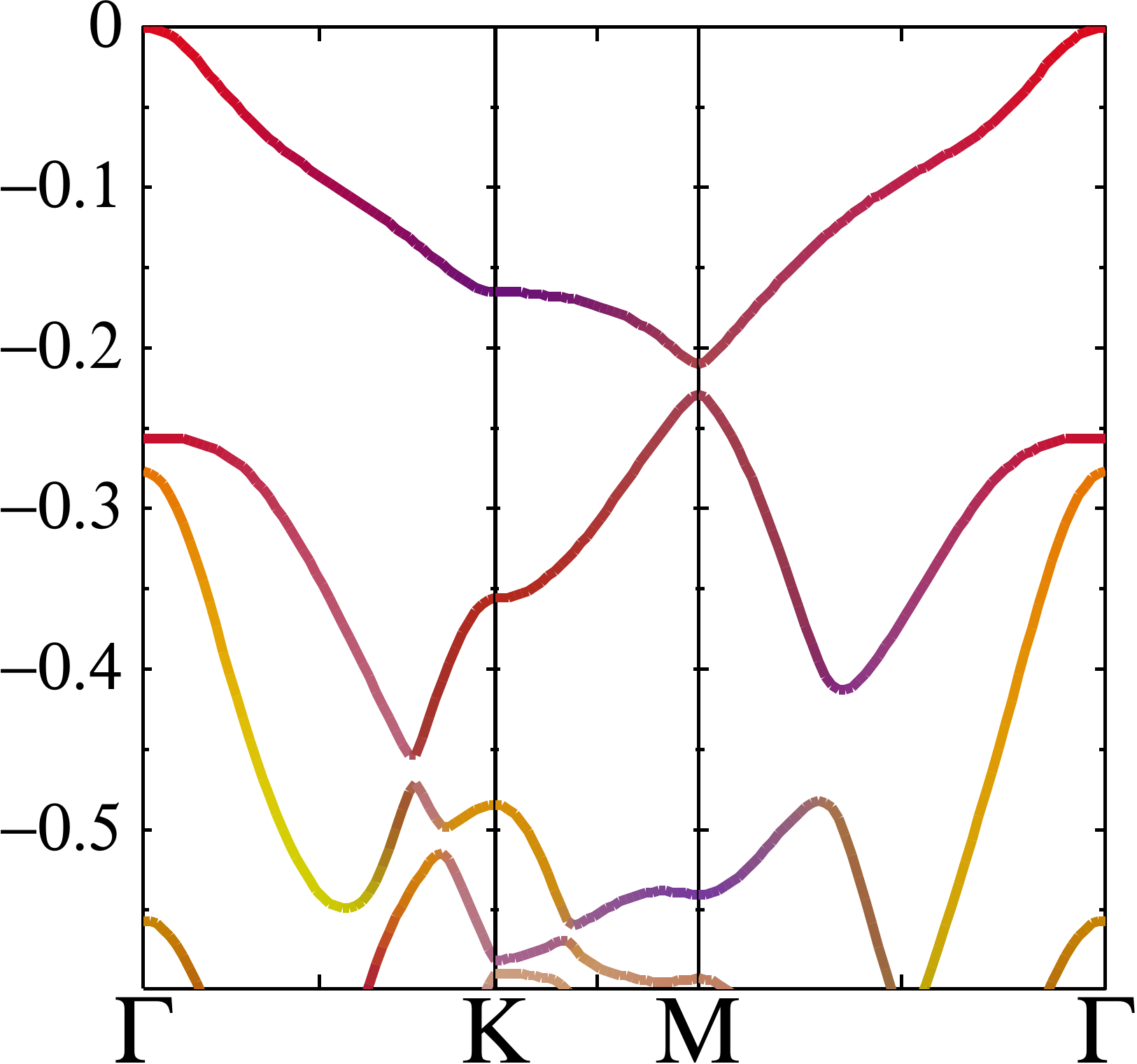}
			\includegraphics[width=0.318\columnwidth, angle=-0]{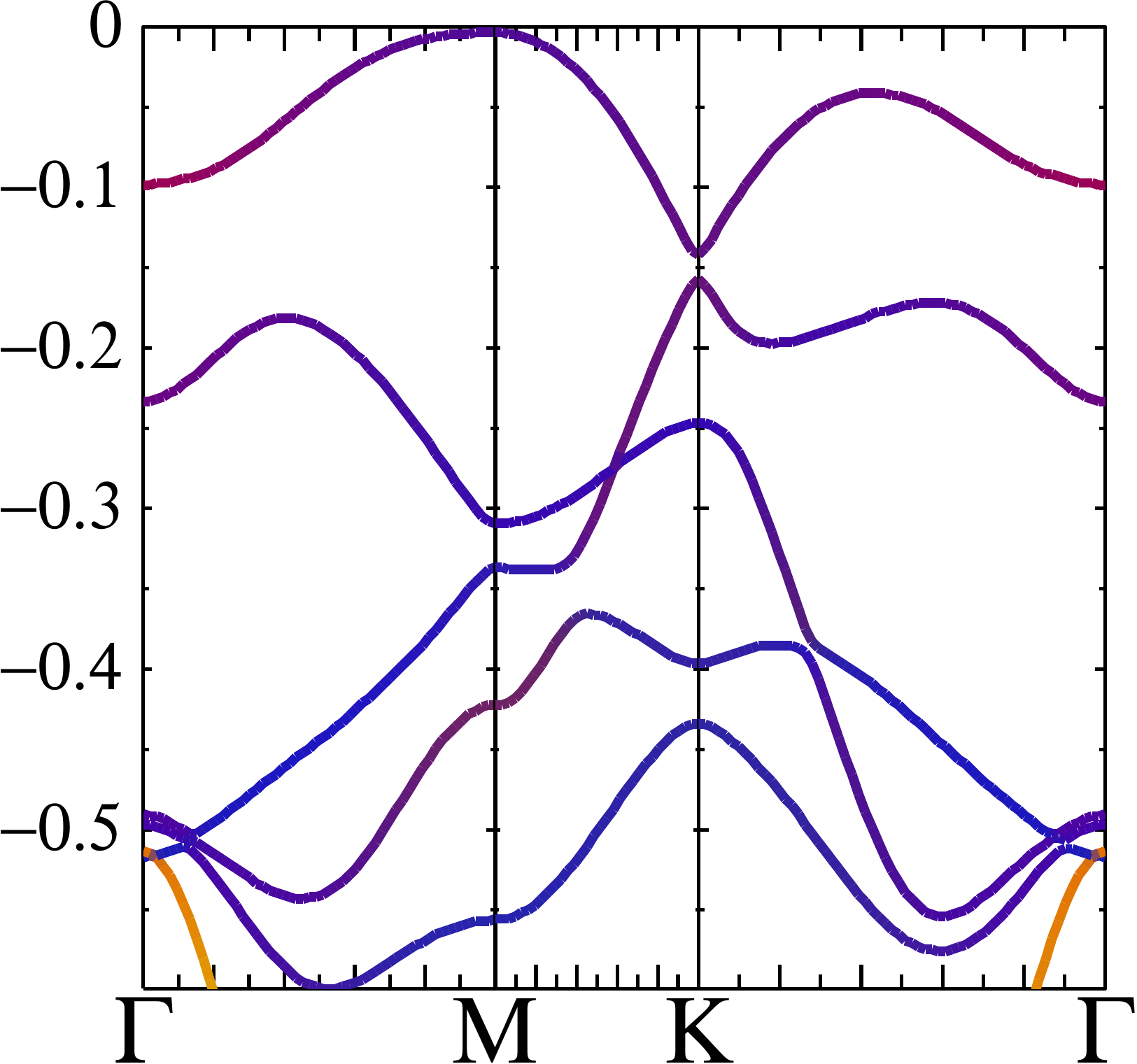}\\
			\includegraphics[width=1.05\columnwidth]{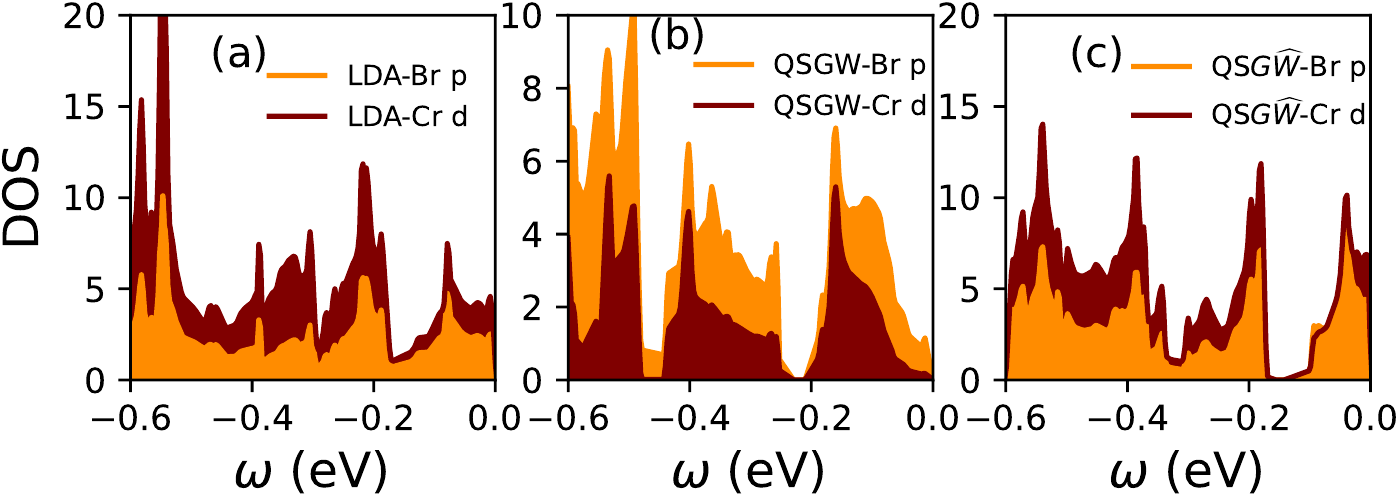}
			\caption{\ce{CrBr3} : From left to right: LDA, QS\emph{GW} and QS$G\widehat{W}$ valence band structures (with spin-orbit coupling) are shown respectively. The colors correspond to Br-$p_{x}+p_{y}$ (red), Br-p$_{z}$ (green), Cr-d (blue). In the bottom panels are shown the density of states projected onto the Cr-d and Br-p states at different levels of the theory.  Structure of the top most valence band is similar within LDA and QS$G\widehat{W}$, but is different in QS\emph{GW}. the QS$G\widehat{W}$ top most valence band is much narrower in comparison to both QS$G\widehat{W}$ and LDA and is split from the rest of the valence band manifold.}
			\label{fig:br}
		\end{center}
	\end{figure}

\begin{figure}
		\begin{center}
			\includegraphics[width=0.34\columnwidth, angle=-0]{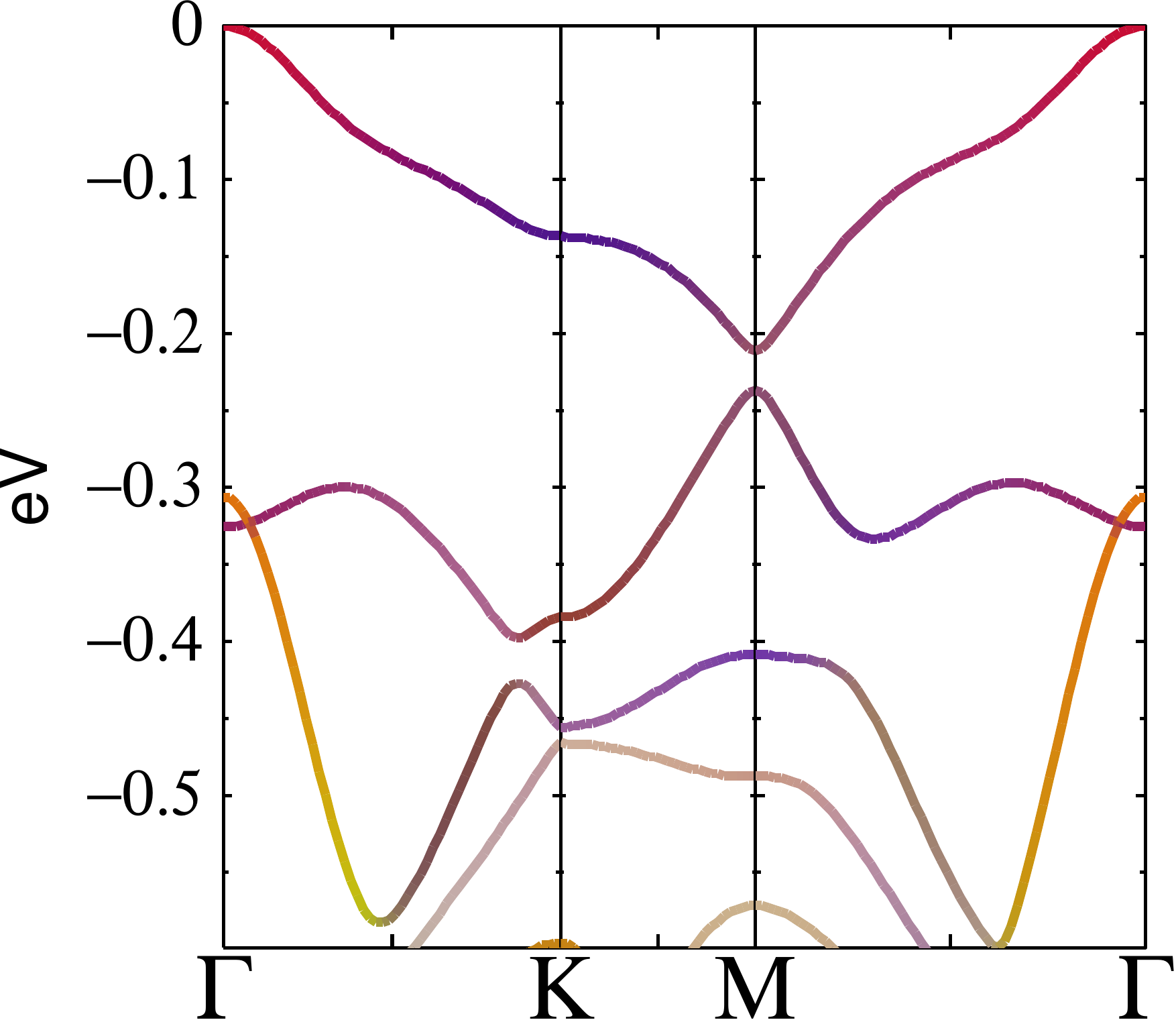}
			\includegraphics[width=0.318\columnwidth, angle=-0]{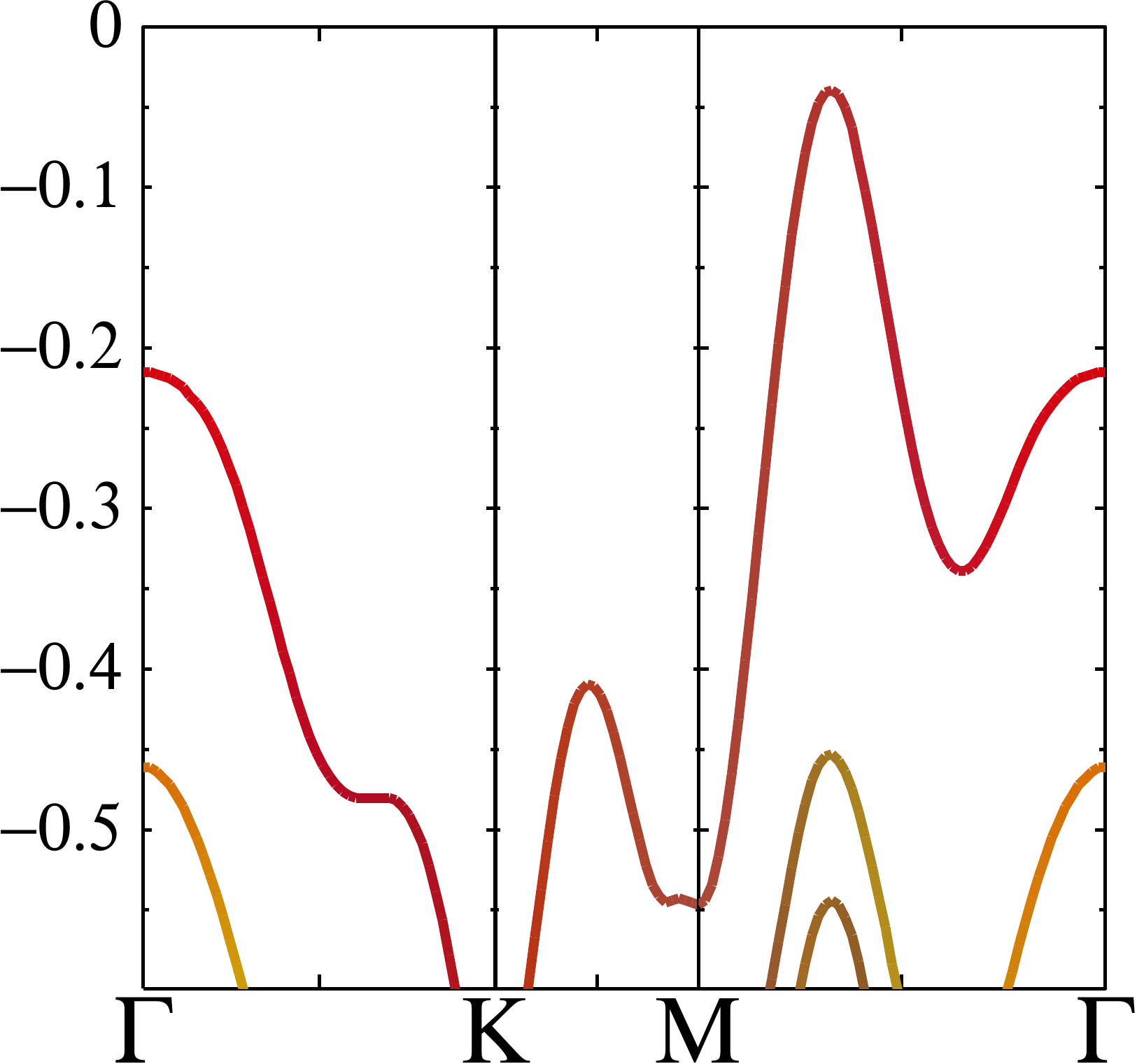}
			\includegraphics[width=0.318\columnwidth, angle=-0]{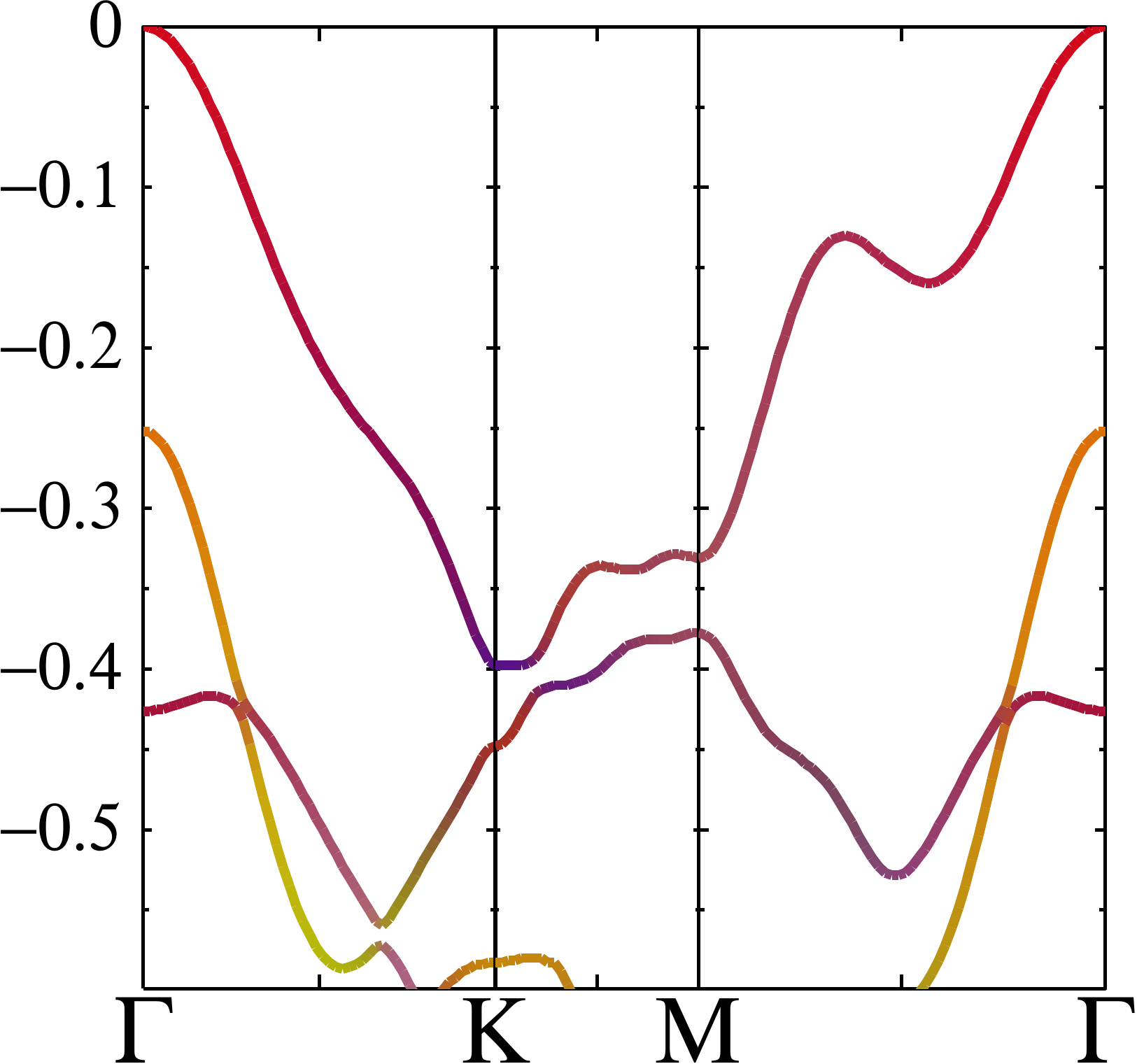}\\
			\includegraphics[width=1.05\columnwidth]{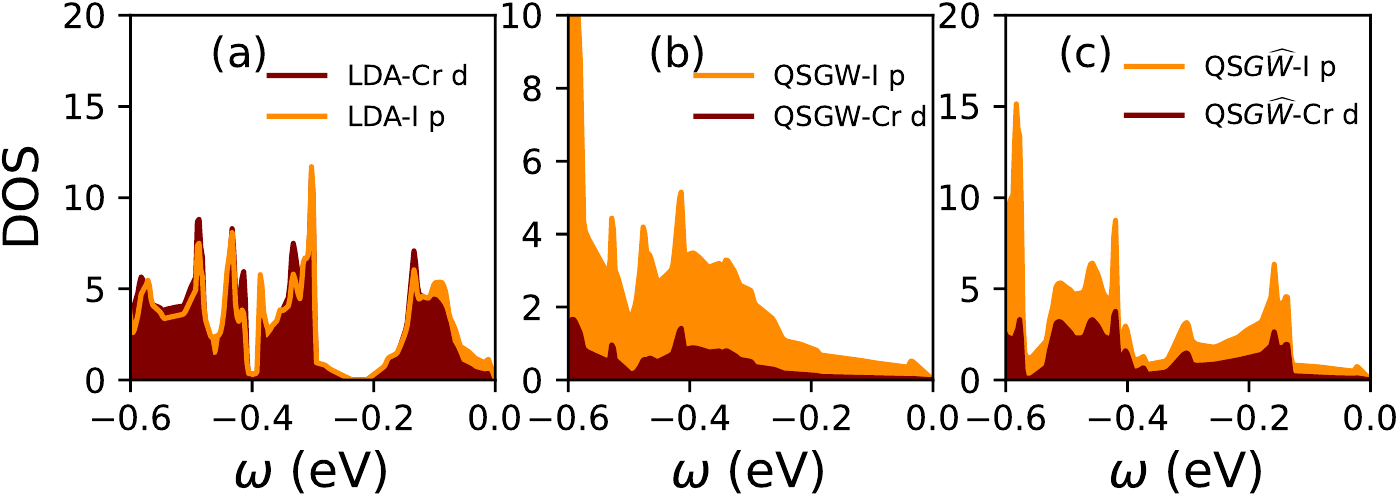}
			\caption{\ce{CrI3} : From left to right: LDA, QS\emph{GW} and QS$G\widehat{W}$ valence band structures (with spin-orbit coupling) are shown
				respectively. The colors correspond to I-$p_{x}+p_{y}$ (red), I-p$_{z}$ (green), Cr-d (blue). In the
				bottom panels are shown the density of states projected onto the Cr-d and I-p states at different
				levels of the theory. Note that in the QS\emph{GW} case, the valence band maximum is at a low-symmetry
				point not on the lines of the figure.}
			\label{fig:i}
		\end{center}
	\end{figure}

	However, as is typical of the LDA, the bandgaps are underestimated, and for \ce{CrX3} the underestimation is severe.
	Accordingly, we study the electronic structure at three different levels of theory: the LDA, the Quasiparticle Self-Consistent
	\emph{GW} approximation (QS\emph{GW}) ~\cite{kotani,pashov}, and an extension of QS\emph{GW} where the RPA approximation
	to the polarizability is extended by adding ladder diagrams (QS$G\widehat{W}$)~\cite{bseoptics,brian}. The electronic dispersions and
	corresponding DOS are shown in Figs.~\ref{fig:cl},\ref{fig:br},\ref{fig:i}, for each level of theory and each material.  In contrast to
	typical \emph{sp} semiconductors, not only the bandgaps but also the valence band dispersions significantly change as the level of theory
	increases.
	
	QS\emph{GW} dramatically enhances the gaps relative to the LDA, as is standard in polar compounds~\cite{kotani}.
	Nevertheless, within the random phase approximation (RPA), it has long been known that $W$ is universally too
	large~\cite{albrecht,rolfing}, and this is reflected in an underestimate of the static dielectric constant
	$\epsilon_\infty$.  Empirically, $\epsilon_\infty$ seems to be underestimated in QS\emph{GW} by a nearly universal
	factor of 0.8, for a wide range of insulators~\cite{chantis,walter}.  Roughly speaking, at low energy $W$ is universally
	too large by a factor 1/0.8 \cite{deguchi} and as a result, QS\emph{GW} bandgaps are slightly
	overestimated~\cite{kotani}.  This can be corrected by extending the RPA to introduce an electron-hole attraction in
	virtual excitations.  These extra (ladder) diagrams are solved by a BSE, and they significantly improve on the optics,
	largely eliminating the discrepancy in $\epsilon_\infty$ \cite{bseoptics}.  When ladders are also added to improve $W$
	in the \emph{GW} cycle ($W{\rightarrow}\widehat{W}$), it significantly improves the one-particle gap as well, as will be
	discussed elsewhere~\cite{brian}. This scenario is played out in \ce{CrX3} compounds: QS\emph{GW} bandgaps are slightly
	larger than QS$G\widehat{W}$ bandgaps, as seen in Table~\ref{tab:gaps}.
	
	Remarkably, the structure of the valence band is very sensitive to the level of theory used, which applies to both, the band energies and wave functions.  Just for \ce{CrCl3} the valence band maximum is independent of the
	theory and is consistently pinned to the M point (Fig.~\ref{fig:cl}).  In the sequence Cl{\textrightarrow}Br{\textrightarrow}I, there
	is an overall tendency for the valence band maximum to shift from the M point to the $\Gamma$ point.  In the
	LDA this transition occurs after Br and I, while in QS\emph{GW} the valence band at $\Gamma$ is above M already for Br.
	QS$G\widehat{W}$ shows the same tendency as QS$G{W}$, but the change is less pronounced and the transition takes place
	between Br and I.  This is a reflection of the softening effects of the ladder diagrams on $W$.  Recent works
	implementing single-shot \emph{GW}, with approximations different from QS\emph{GW}, also finds the valence band maximum
	in \ce{CrI3} at $\Gamma$~\cite{molina,louie}, which also seems to be confirmed by a recent ARPES study~\cite{kundu}.
	
		\begin{table}[t]
		\footnotesize
		\begin{tabular}{c@{\hskip 3pt}|cc@{\hskip 3pt}|@{\hskip 3pt}cc@{\hskip 3pt}|@{\hskip 3pt}cc@{\hskip 3pt}}
			theory & \multicolumn{2}{c}{\ce{CrCl3}} & \multicolumn{2}{c}{\ce{CrBr3}}  & \multicolumn{2}{c}{\ce{CrI3}} \cr
			& $m_x$&  $m_y$  & $m_x$ & $m_y$ & $m_x$& $m_y$  \cr
			\hline
			\vbox{\vskip 12pt}
			DFT    & 1.9  &  3.6    & 2.0        & 5.2       & 1.2  & 1.2 \cr
			QSGW   & 2.3  &  5.5    & $\infty$   & -1.3      & 1.4  & 1.4 \cr
			QS$G\widehat{W}$    & 2.1  &  4.2    & 3.2        & $\infty$  & 0.57 & 0.57 \cr
		\end{tabular}
		\caption{Effective masses $m^*/m_0$ at the M point (as shown in Fig.~\ref{fig:struc}) for CrCl\textsubscript{3} and CrBr\textsubscript{3}, and at the
			$\Gamma$ point for CrI\textsubscript{3}, for three levels of approximation.  These \emph{k} points correspond to
			the valence band maximum except for CrBr\textsubscript{3} in the QS\emph{GW} approximation (see Fig.~\ref{fig:i}). $m_x$ and $m_y$ correspond to orientations perpendicular and parallel to the $\Gamma$-M line, respectively.  $\infty$ is a shorthand for an effective mass larger than 10$m_0$.
			}
		\label{tab:masses}
	\end{table}
	
	\begin{figure}
		\begin{center}
			\includegraphics[width=0.47\columnwidth]{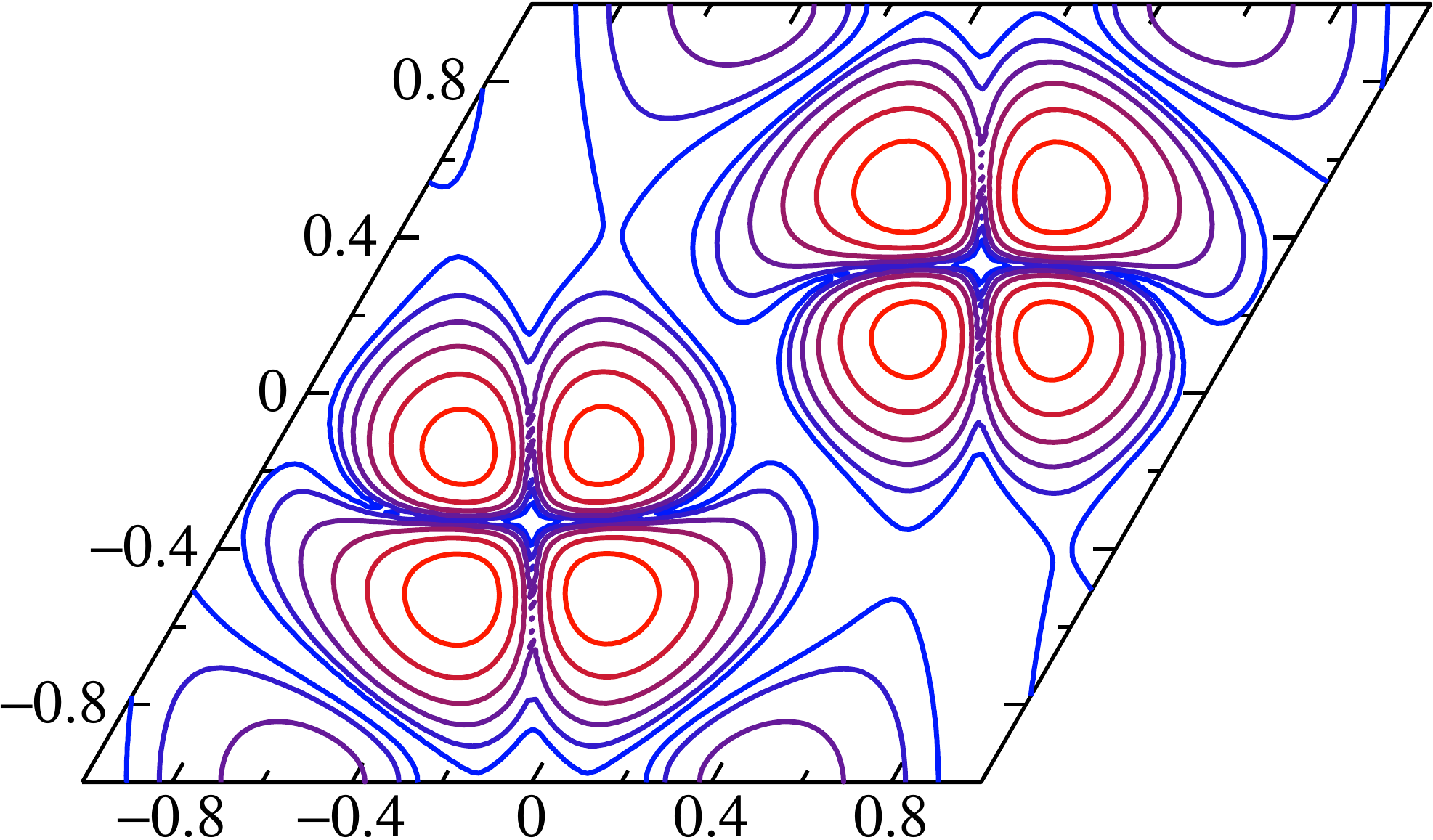}
			\includegraphics[width=0.47\columnwidth]{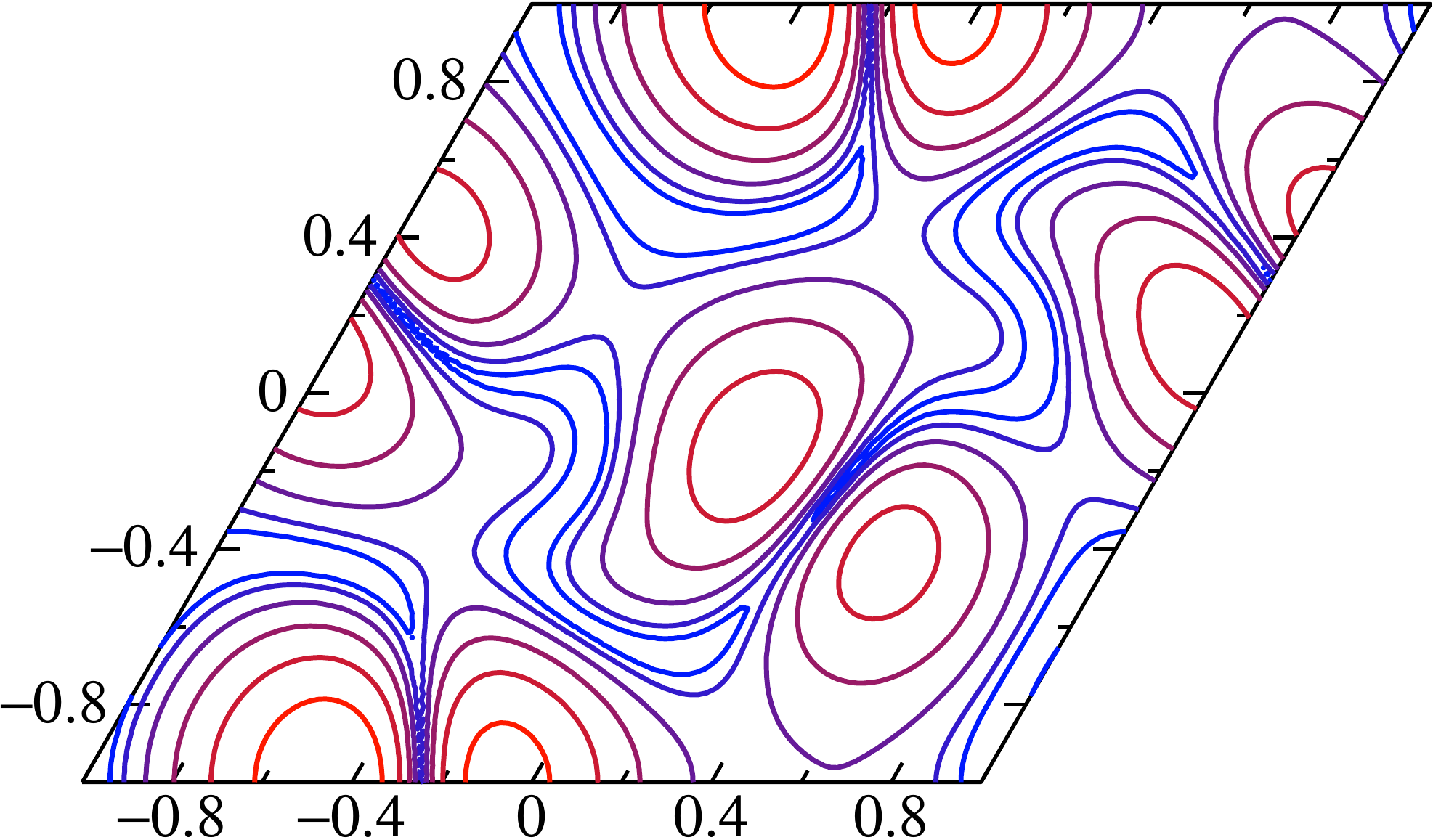}
			\includegraphics[width=0.47\columnwidth]{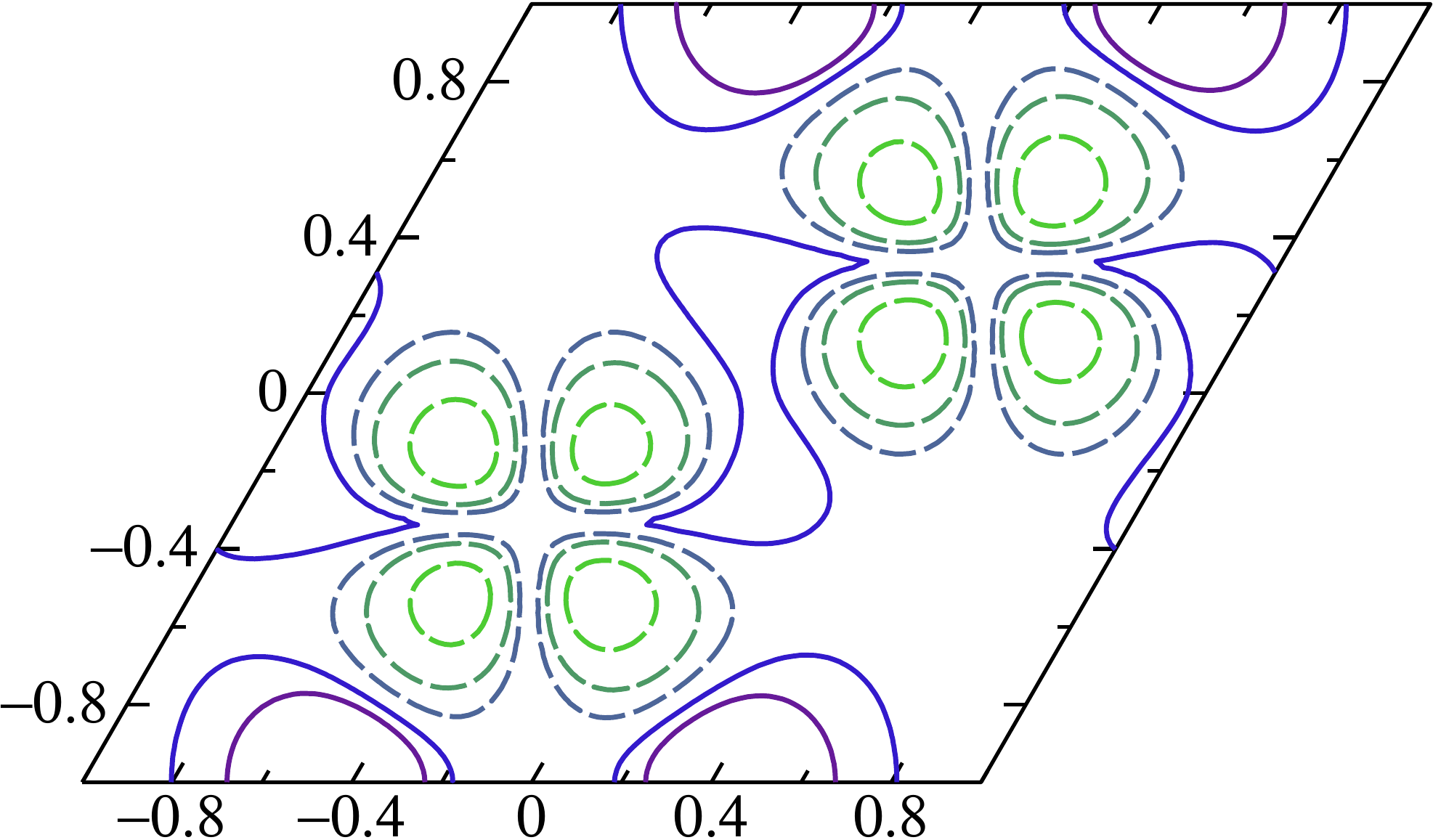}
			\includegraphics[width=0.47\columnwidth]{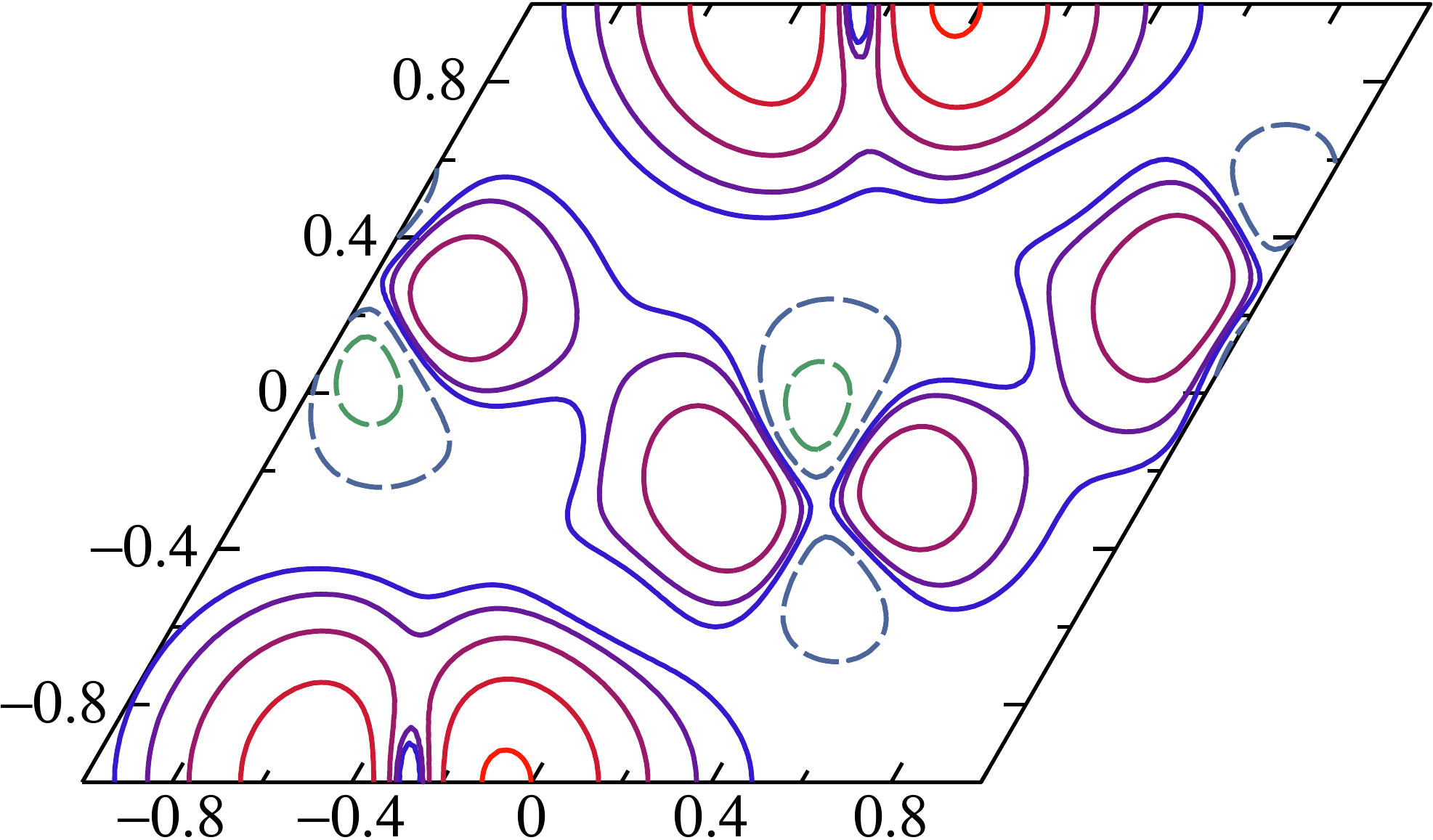}
			\includegraphics[width=0.47\columnwidth]{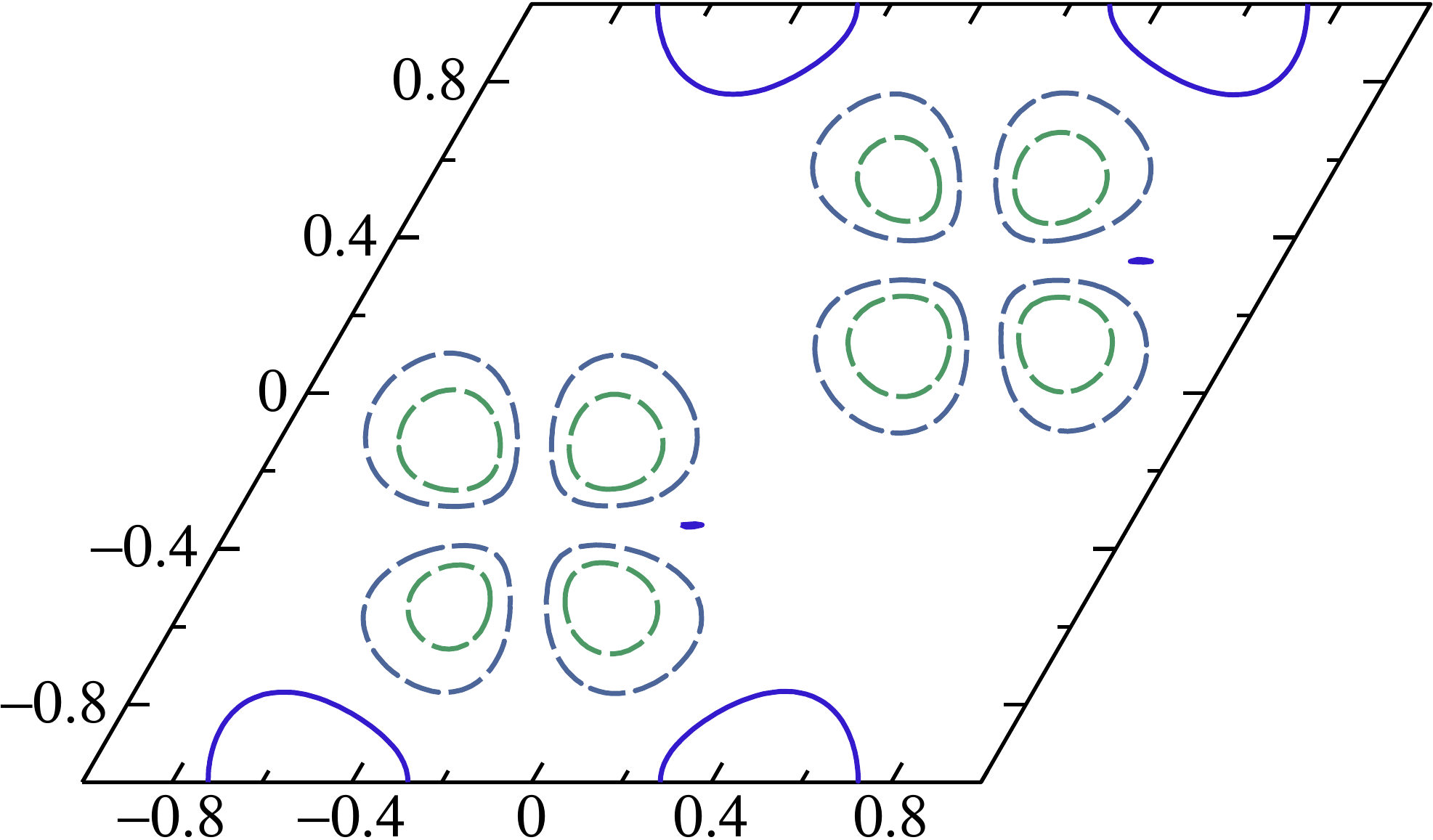}
			\includegraphics[width=0.47\columnwidth]{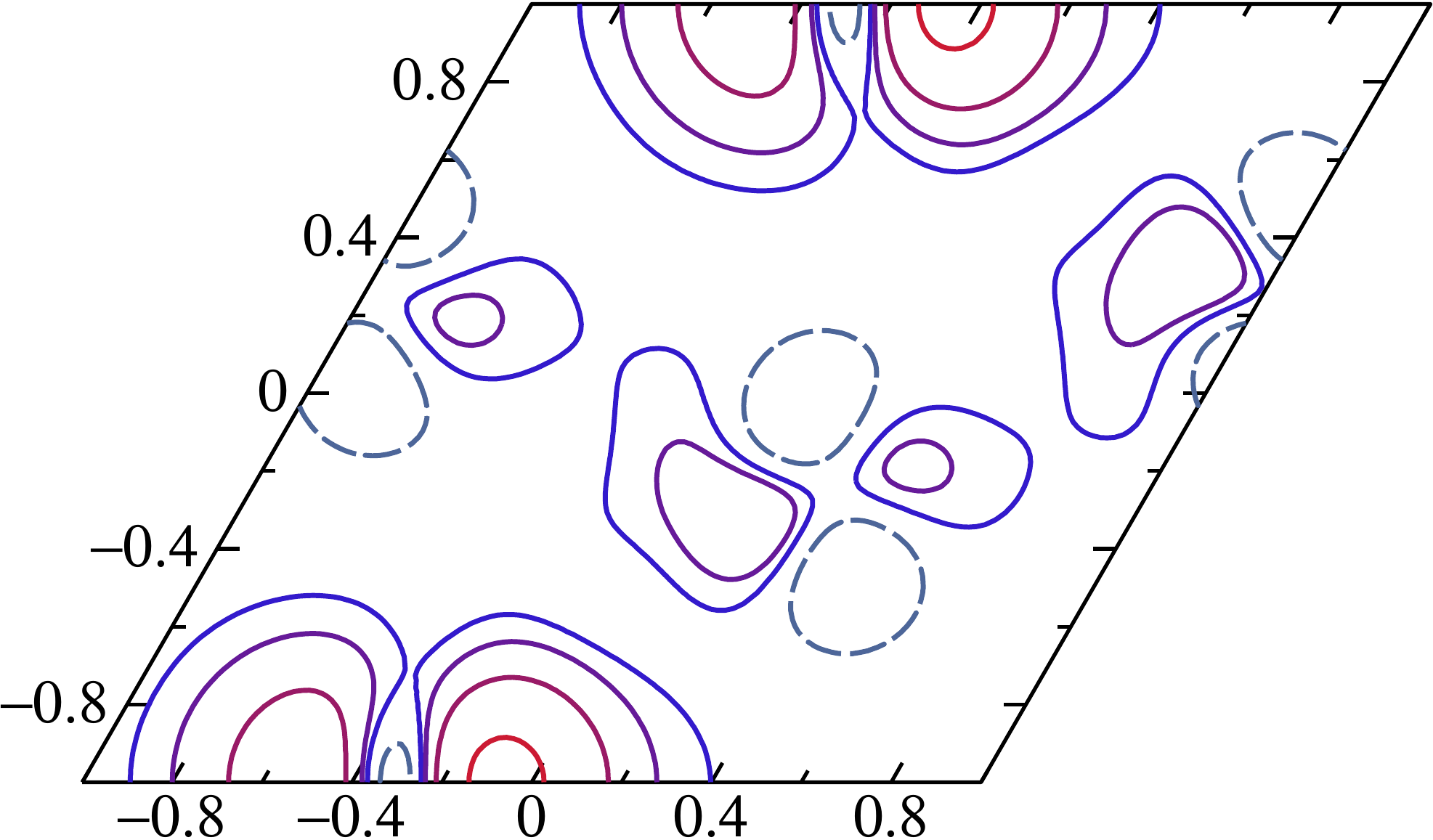}
			\caption{\ce{CrBr3} : Square of wave function $\psi$, in real space, of the highest valence band state at the M point.
				All the left panels pass through a Cr plane and right panels pass through a Br plane.  Top panels
				display constant-amplitude contours for DFT eigenfunctions.  Contours are taken in half-decade
				increments in $|\psi|^2$, with a factor of 300 between highest contour (red) and lowest (blue).  In
				the Cr plane, the atomic $d_{xy}$ character centered at Cr nuclei stand in sharp relief; in the
				Br plane the Br atomic \emph{p} character is evident.  Middle panels show the change in
				$|\psi|^2$ passing from DFT to QS\emph{GW} eigenfunctions; bottom panels show the corresponding
				change passing from DFT to QS$G\widehat{W}$ eigenfunctions.  In the bottom four panels, blue{\textrightarrow}red
				has a similar meaning as in the top panels (increasing positive ${\delta}|\psi|^2$), while contours
				of negative ${\delta}|\psi|^2$ are depicted by increasing strength in the change
				blue{\textrightarrow}green.
								}
				
			\label{fig:density}
		\end{center}
	\end{figure}
	
	While the band maximum is roughly similar across different levels of theory, there are sharp differences in the overall electronic dispersions and wave functions.
	Even though the QS$G\widehat{W}$ band structure more closely resembles DFT than QS\emph{GW}, the eigenfunctions do not.  
	This can be
	seen by inspecting the square of the wave function, $|\psi^2|$, corresponding to the highest-lying state at
	the M point (Fig.~\ref{fig:density} The density is plotted in real space, and the abscissa and ordinate are defined by the the inverse transpose of the 2$\times$2 matrix composed of $\mathbf{b_1}$ and $\mathbf{b_2}$ of Fig.~\ref{fig:struc}.  Throughout this paper $x$ and $y$ are defined by aligning $\mathbf{b_2}$ parallel to $y$.  In this notation the M point is on the $\mathbf{b_2}$ line, or the $y$ axis.  Contour plots in two planes are shown, passing through Cr and Br, respectively.
	At the LDA level (top panels), the wave function resembles an atomic $d_{xy}$ state centered at each Cr nucleus.  In the
	Br plane some Br \emph{p} character is evident, and the bond is partially directed along \emph{x}.  The middle panels
	depict the change in $|\psi^2|$ when passing from DFT to QS\emph{GW}.  Two effects are prominent: first there is a
	transfer of spectral weight from Cr to Br (mostly green contours on Cr, red on Br) as also noted in
	Table~\ref{tab:gaps}.  Second, the bonding becomes more directional, forming one-dimensional chains along $x$.  This
	reflects an enhancement of the Cr-Cr coupling mediated through the Br.  It is especially apparent in the Br plane, but
	it is also reflected in the asymmetry between the $x$ and $y$ directions in the Cr plane.  The bottom two panels show the change in
	$|\psi^2|$ when passing from DFT to QS$G\widehat{W}$.  The effect is similar to DFT{\textrightarrow}\emph{GW}, but the magnitude of
	the change is softened.
	
	\begin{figure}
		\begin{center}

			      \includegraphics[width=\columnwidth]{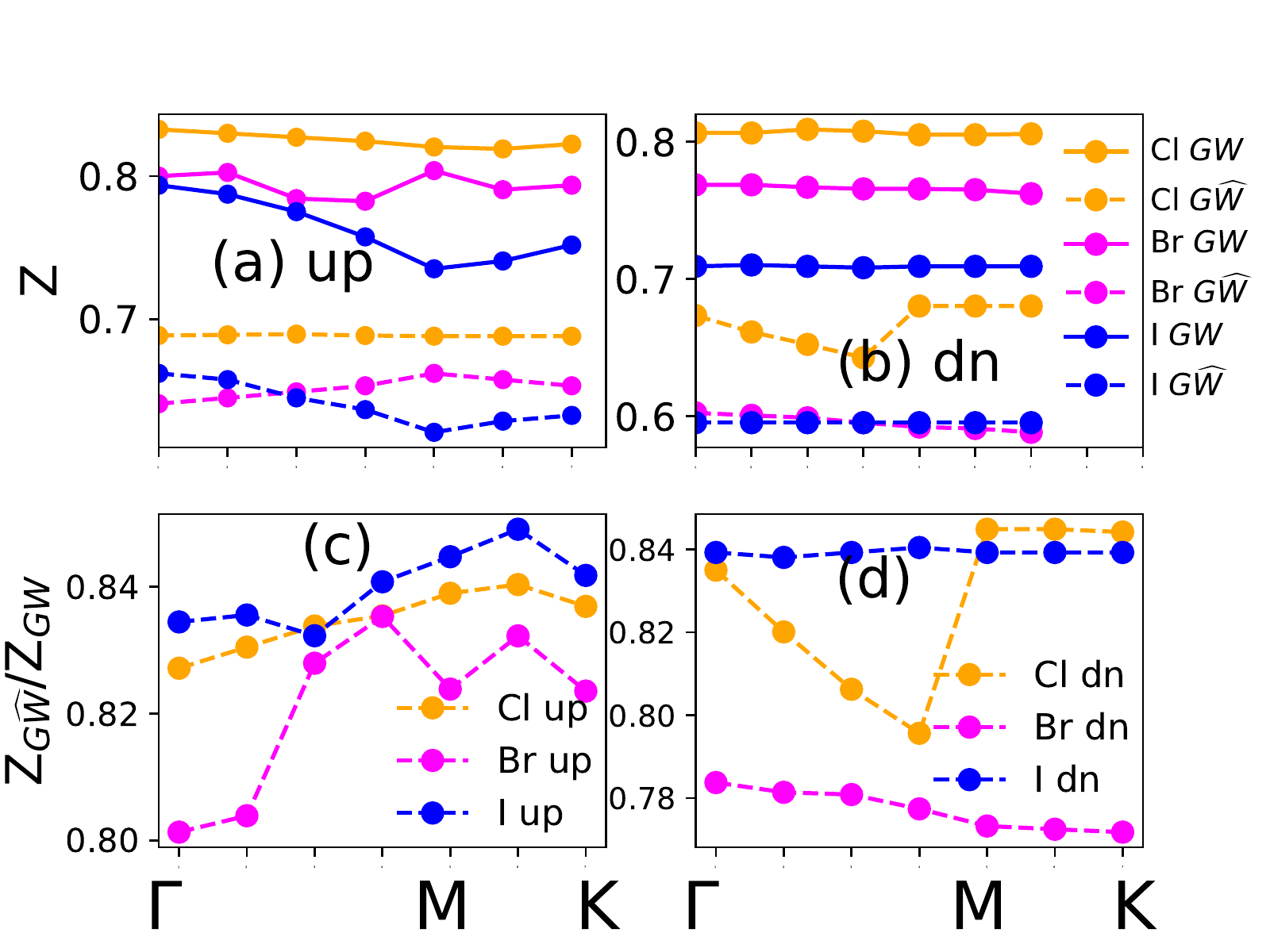}
			\caption{\ce{CrX3} : Real part of $\Sigma(k,\omega)$ is analyzed to extract the quasi-particle renormalization factor Z$_{k}$ from QS\emph{GW}  and QS$G\widehat{W}$.  (a) and (b) show the weak momentum dependence of the Z$_{k}$ for the top most valence band at the quasi-particle peaks at the $k$ points chose along the high symmetry directions of the first Brillouin zone for the up and down spin sectors respectively. (c) and (d) show the relative suppression of the Z$_{k}$ factor in QS$G\widehat{W}$, compared to QS\emph{GW} for the up and down spin sectors respectively. }
			\label{fig:sigma}
		\end{center}
	\end{figure}

	The directionality in the wave function is also reflected in strong anisotropy in the valence band mass at the point M,
	particularly in CrBr\textsubscript{3} (see Table.~\ref{tab:masses}).  By symmetry, there is no anisotropy at the $\Gamma$ point, but at M it becomes
	quite pronounced at the highest level of theory. 
	
	Finally, we analyze the dynamic and momentum dependent self energies $\Sigma(k,\omega)$ from QS\emph{GW} and QS$G\widehat{W}$ to
	further understand the changes in the valence band structure at different levels of theory. We observe that the $\Sigma$
	of the top most valence band has very week dependence on momentum. The momentum dependence is even weaker in the down
	spin channel, that is unoccupied. For the up spin channel the momentum dispersion is very similar both in QS\emph{GW}
	and QS$G\widehat{W}$. We extract the quasi-particle renormalization ($Z_{k}$) factors from $\Sigma(k,\omega)$ at the quasi-particle
	energies for the top most valence band. We observe that Z$_{k}$ reduces by $\sim$20\% within QS$G\widehat{W}$ in comparison to
	QS\emph{GW}. This suggests that the quasi-particles become further localized at the QS$G\widehat{W}$ level, in comparison to the
	QS\emph{GW}. This goes along with charge density that is weakly put back on the atoms at the QS$G\widehat{W}$ level.\\
	
	\noindent\emph{Conclusions}
	
	We analyze in detail the electronic band structure of CrX\textsubscript{3} within different levels of an
	\emph{ab initio} theory.  The results were interpreted in terms of a simplified tight-binding model to elucidate the
	trends in Cl{\textrightarrow}Br{\textrightarrow}I, in particular the bandgap and the orbital character of the valence
	band.  Many-body effects both enhance the bandgap, and make the valence band eigenfunctions more directional.  We also
	showed that addition of ladder diagrams to improve \emph{W} increases the screening, thus softening the many body
	corrections to DFT. Further we quantify the momentum dependence of the self-energies at different levels of the theory and show explicitly how excitonic correlations lead to re-normalization of the electronic bands and localization of charges. Summarily, we show how a starting point independent implementation of \emph{GW} and \emph{BSE} leads to changes in electronic band energies and wavefunctions via complicated interplay between charge and self-energy self-consistencies in CrX$_{3}$.\\
	
	\noindent\emph{Acknowledgment}
	
	MIK, ANR and SA are supported by the ERC Synery Grant, project 854843 FASTCORR (Ultrafast dynamics of
	correlated electrons in solids). MvS and DP are supported by the Simons Many-Electron Collaboration.  We acknowledge PRACE for awarding us access to
	Irene-Rome hosted by TGCC, France and Juwels Booster and Clusters, Germany; STFC Scientific Computing Department's SCARF cluster, Cambridge Tier-2 system operated by
	the University of Cambridge Research Computing Service (www.hpc.cam.ac.uk) funded by EPSRC Tier-2 capital grant
	EP/P020259/1.

\appendix

\section{Numerical Details}

    Single particle calculations (DFT, and energy band calculations with the static quasiparticlized QS\emph{GW} self-energy
	$\Sigma^{0}(k)$) were performed on a 16$\times$16$\times$1 \emph{k}-mesh while the (relatively smooth) dynamical self-energy
	$\Sigma(k)$ was constructed using a 6$\times$6$\times$1 \emph{k}-mesh and $\Sigma^{0}$(k) extracted from it.  For each
	iteration in the QS\emph{GW} self-consistency cycle, the charge density was made self-consistent.  The QS\emph{GW} cycle
	was iterated until the RMS change in $\Sigma^{0}$ reached 10$^{-5}$\,Ry.  Thus the calculation was self-consistent in
	both $\Sigma^{0}(k)$ and the density.  Numerous checks were made to verify that the self-consistent $\Sigma^{0}(k)$ was
	independent of starting point, for both QS$GW$ and QS$G\widehat{W}$ calculations; e.g. using LDA or Hartee-Fock self-energy as the initial self energy for QS\emph{GW} and using LDA or QS\emph{GW} as the initial
	self-energy	for QS$G\widehat{W}$.
	
	For the present work,  the electron-hole two-particle correlations are incorporated within a self-consistent ladder-BSE implementation~\cite{bseoptics,brian} with Tamm-Dancoff
	approximation~\cite{tamm1,tamm2}. The effective interaction \emph{W} is calculated with ladder-BSE corrections and the self energy, using a static vertex in the
	BSE.
	\emph{G} and \emph{W} are updated iteratively until all of them converge and this is what we call QS$G\widehat{W}$.  Ladders increase the screening of \emph{W},
	reducing the gap besides softening the LDA\textrightarrow{QS\emph{GW}} corrections noted for the valence bands. 

	For all
	materials, we checked the convergence in the QS$G\widehat{W}$ band gap by increasing the size of the two-particle Hamiltonian. We
	increase the number of valence and conduction states that are included in the two-particle Hamiltonian. We observe that
	for all materials the QS$G\widehat{W}$ band gap stops changing once 24 valence and 24 conduction states are included in the
	two-particle Hamiltonian. While the gap is most sensitive to the number of valence states, 14 conducting states produces
	results within 2\% error of the converged results from 24 conduction states.
	
	In Tab.~\ref{table3} we list the Cr-d occupancies for different materials at different levels of the theory. 
	
    \begin{table}
    	\footnotesize
    	\begin{tabular}{ccccccccccc}
    		\hline
    		variants & \ce{CrCl3} & \ce{CrBr3} & \ce{CrI3}  \\
    		\hline
    		DFT & 4.23  & 4.44 & 4.66\\
    		QS\emph{GW}  & 4.08  &  4.3 &  4.66 \\
    		BSE   & 4.11 & 4.35 &  4.64  \\
    	\end{tabular}
    	\caption{Shown are the Cr-d occupancies. }
    	\label{table3}
    \end{table}	
	
\section{Vacuum Distance Scaling}

	\begin{figure}
		\includegraphics[width=0.8\columnwidth]{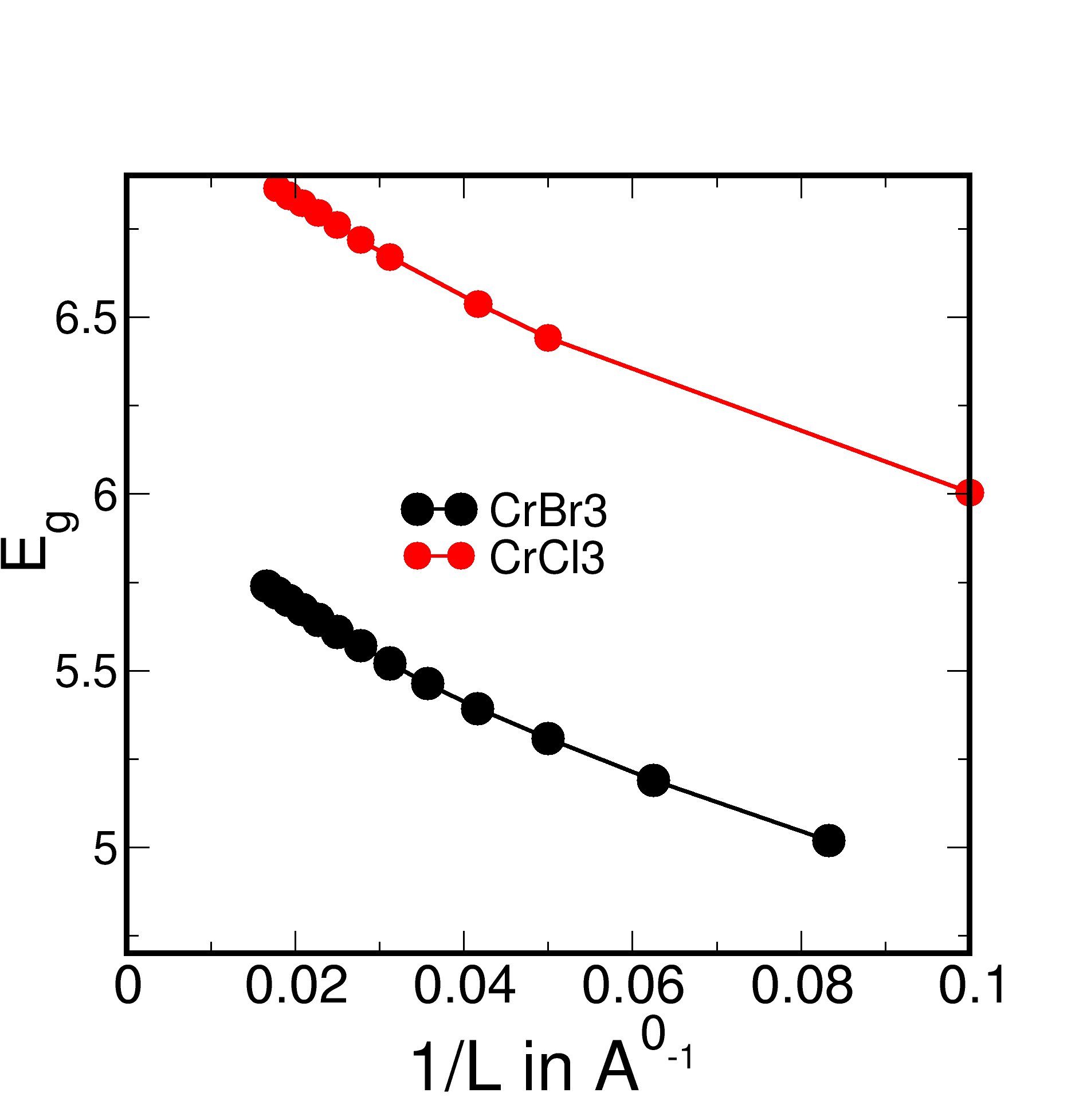}
		\caption{\ce{CrBr3 and CrCl3} : Scaling of the one particle band gap from QSGW with vacuum size \emph{L}.}
		\label{fig:scaling}
	\end{figure}

	Periodic boundary conditions were used in all directions, introducing an unwanted coupling between \ce{CrBr3} slabs.
	To minimize this coupling a vacuum of length \emph{L} was inserted between slabs, and \emph{L} was varied.
	
	As is well known, the
	QS\emph{GW} is known to fix the infamous `gap problem'~\cite{ferdi} in insulators, we observe that the band gap increases
	significantly in all three compounds within QS\emph{GW}; 6.87 eV in Cl, 5.73 eV in Br and 3.25 eV in I (see
	~\ref{tab:gaps}). We change the vacuum length from 10 \AA\, to 80 \AA, and observe the scaling of the band gap with
	vacuum size (L). We observe an almost perfect 1/L scaling (see Fig.~\ref{fig:scaling}) of the gap as noted earlier in a
	separate work on V$_{2}$O$_{5}$ ~\cite{bhandari}. This also allows us to check the dielectric constant
	($\epsilon_{\infty}$) and its vacuum correction. In the limit of a purely free standing monolayer all three directional
	components of the macroscopic dielectric response in the static limit approaches 1, suggesting the absence of
	screening. We use this vacuum length (60 \AA) for the rest of the discussions in the present work.
	
\section{Full Band Structures}

\begin{figure*}
			\begin{center}
	\includegraphics[width=0.34\textwidth, angle=-0]{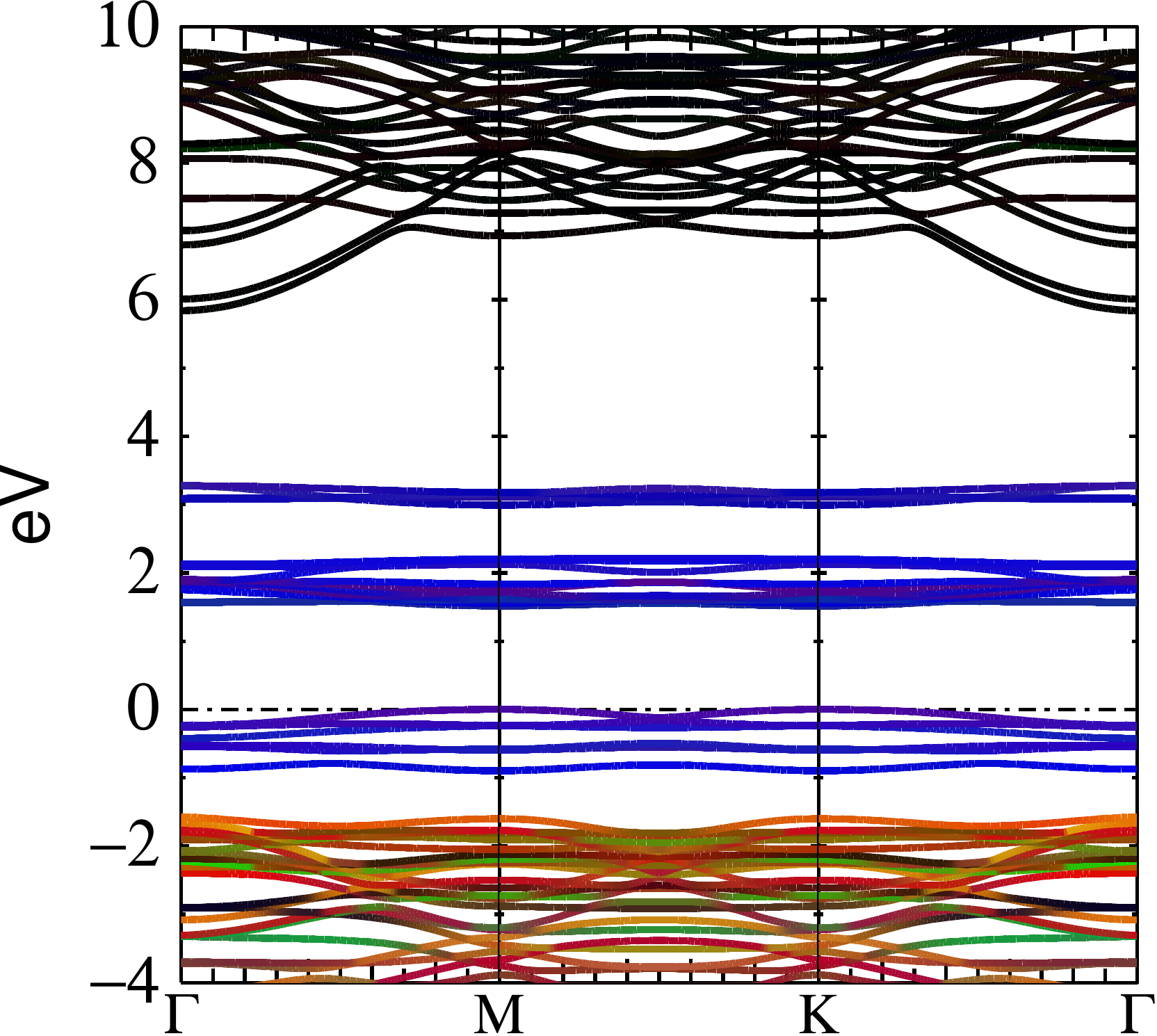}
	\includegraphics[width=0.32\textwidth, angle=-0]{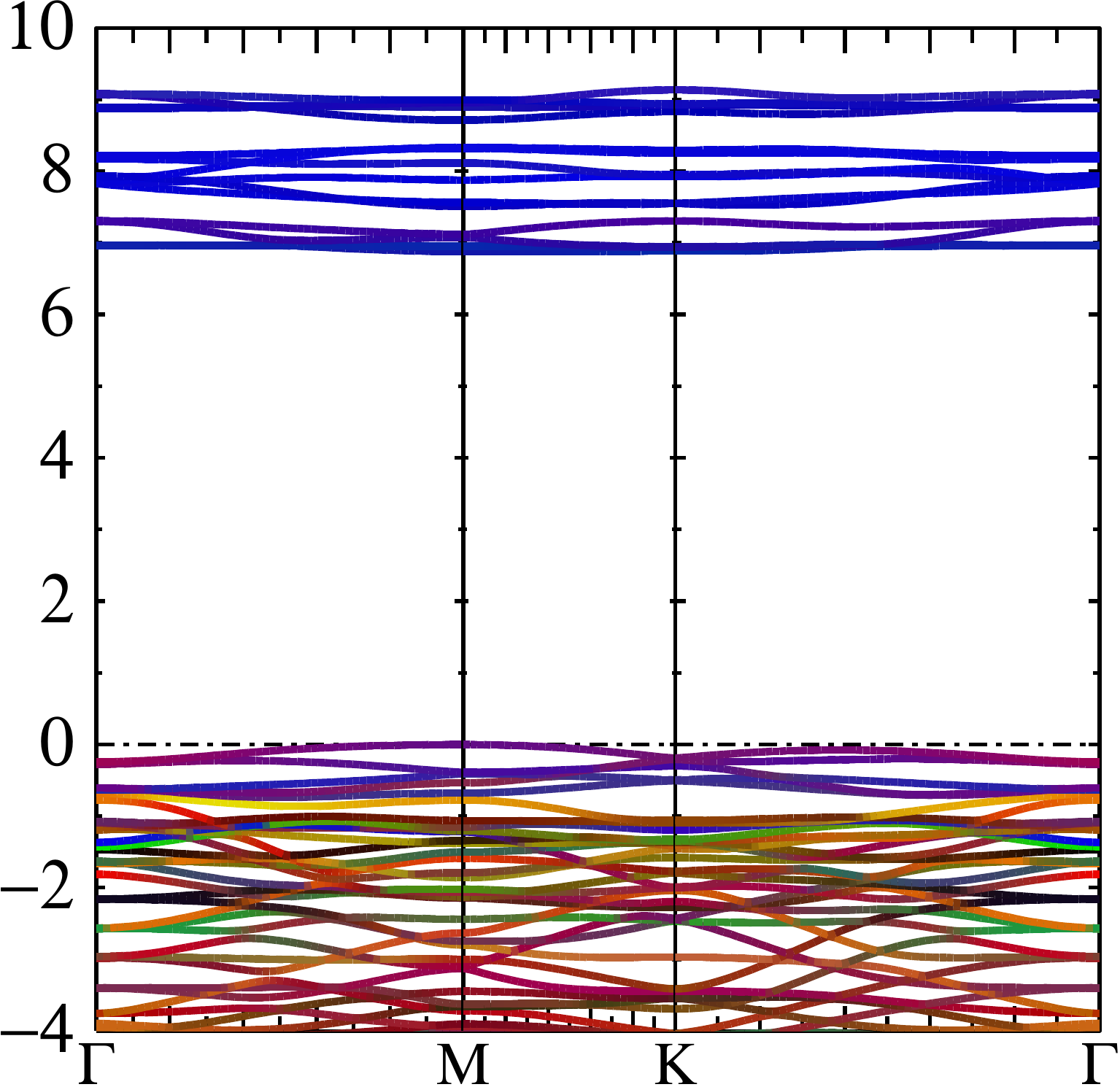}
	\includegraphics[width=0.32\textwidth, angle=-0]{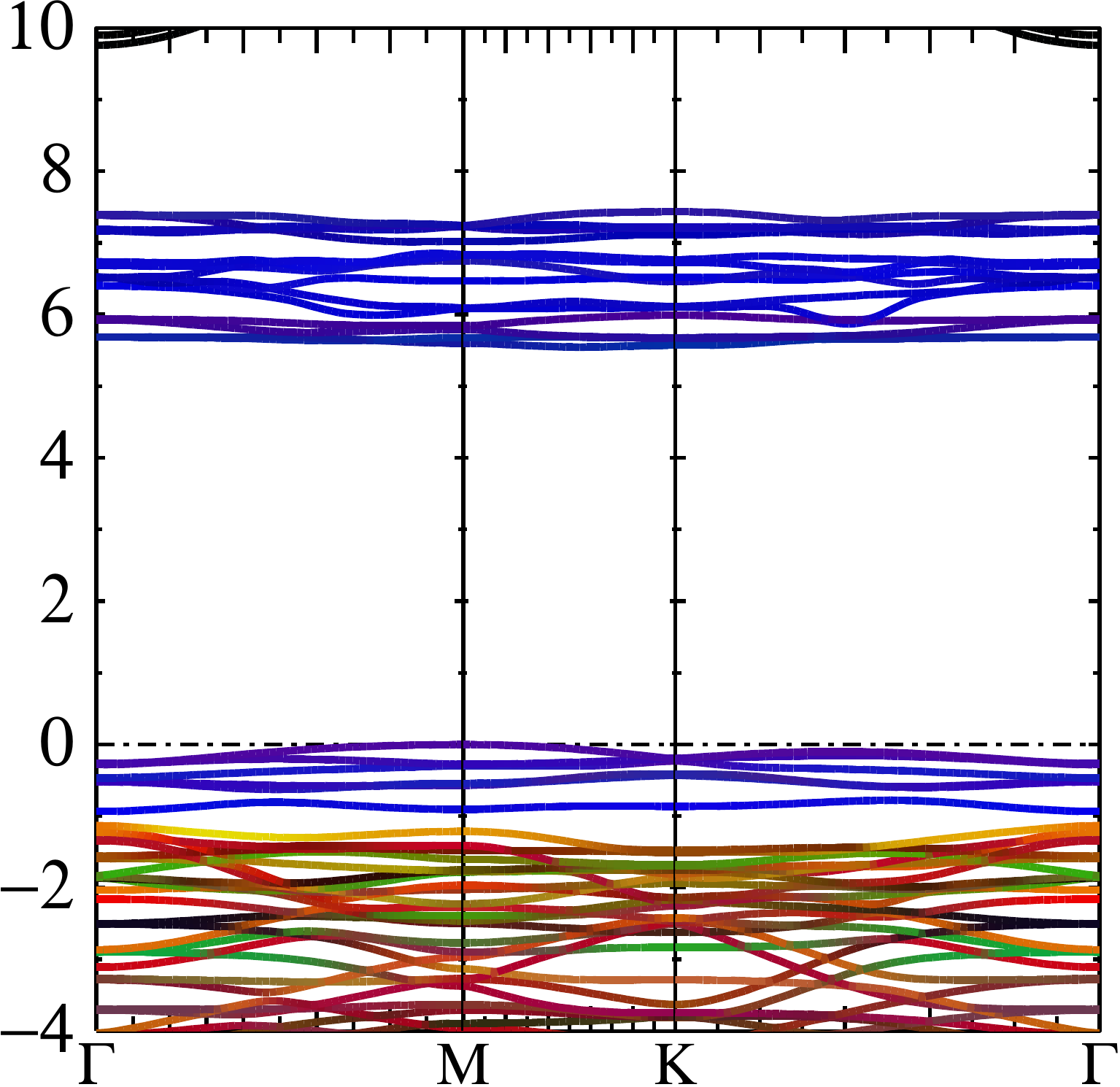}\\
	\includegraphics[width=0.34\textwidth, angle=-0]{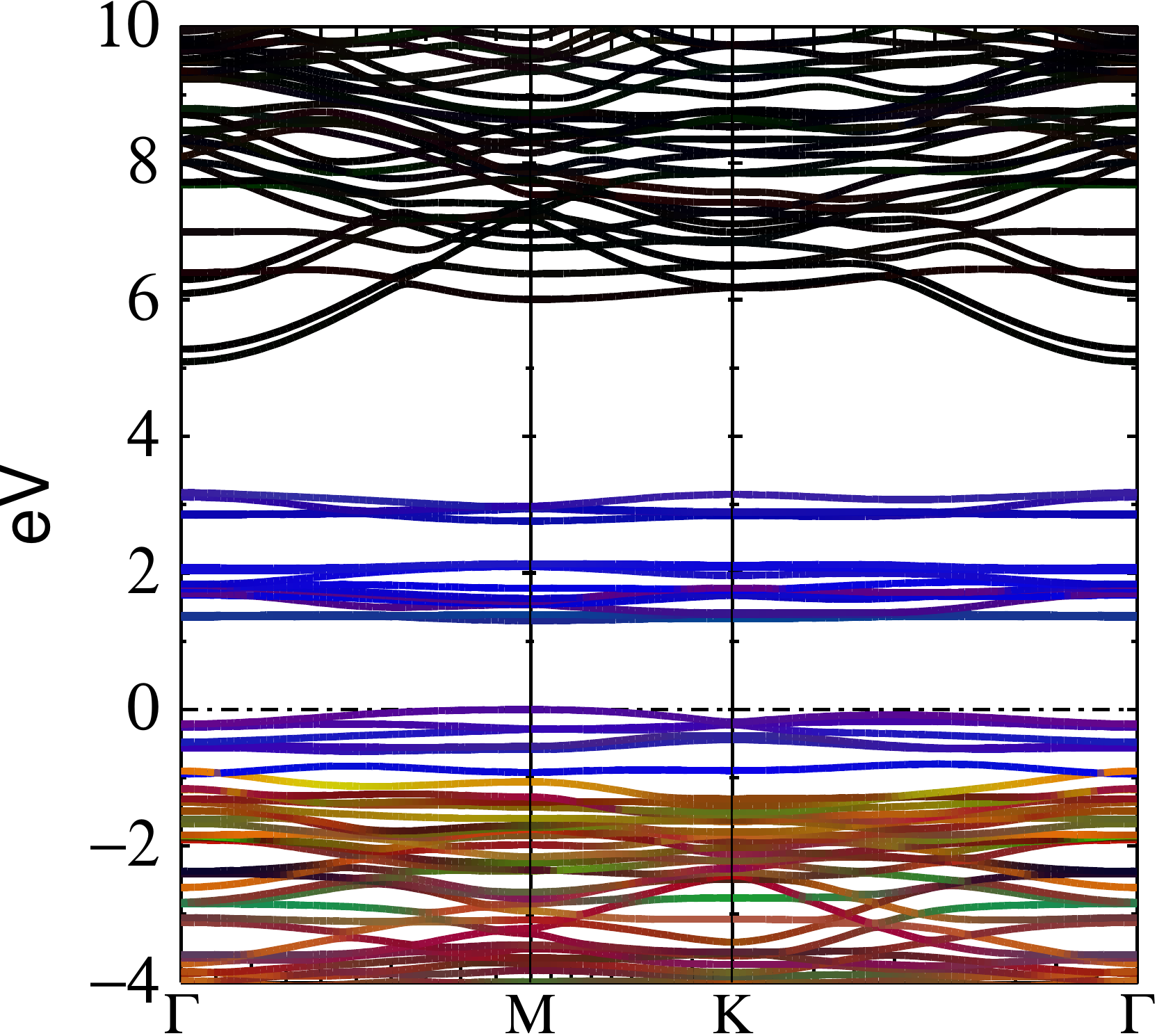}
	\includegraphics[width=0.32\textwidth, angle=-0]{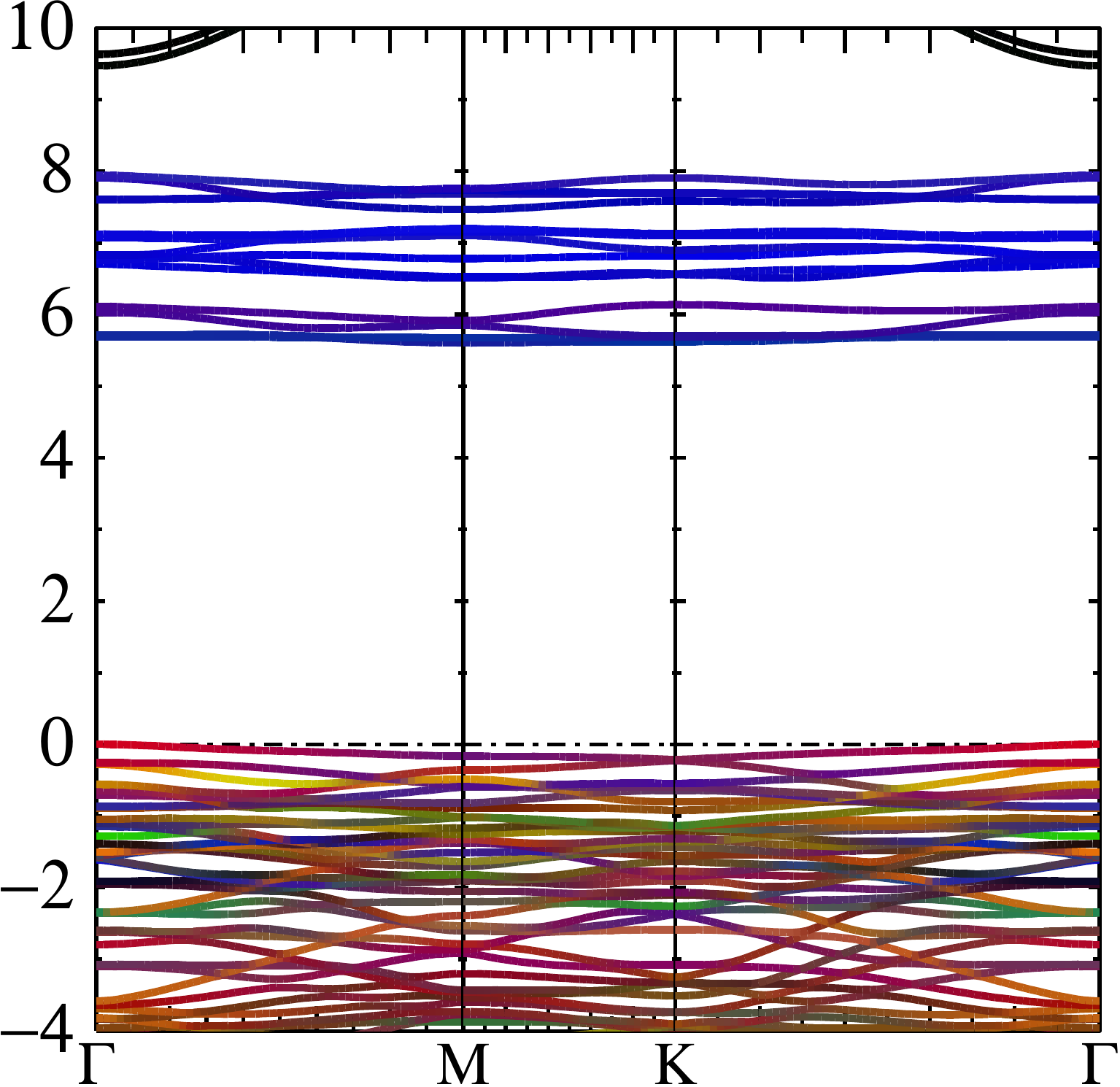}
	\includegraphics[width=0.32\textwidth, angle=-0]{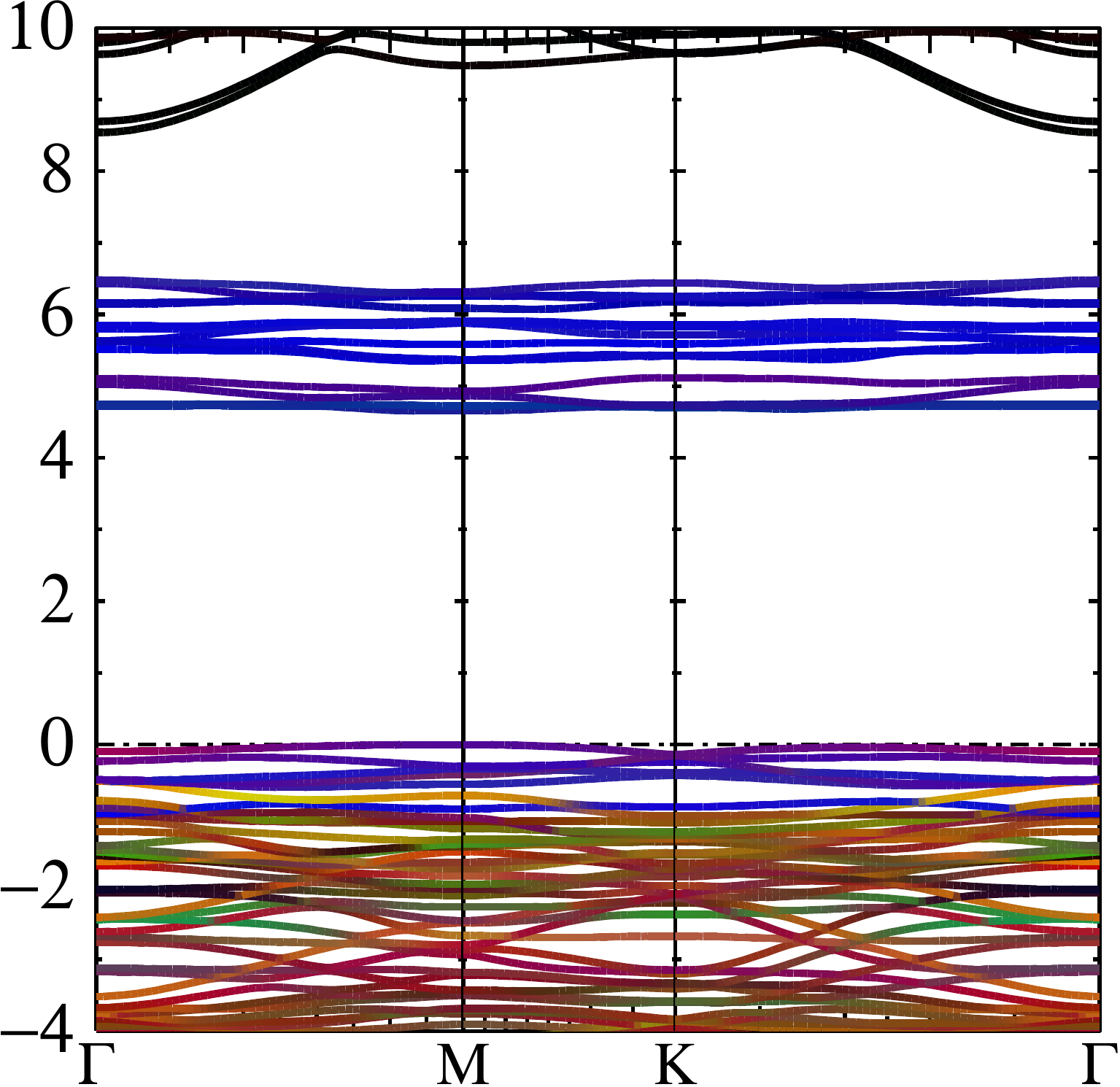}\\
	\includegraphics[width=0.34\textwidth, angle=-0]{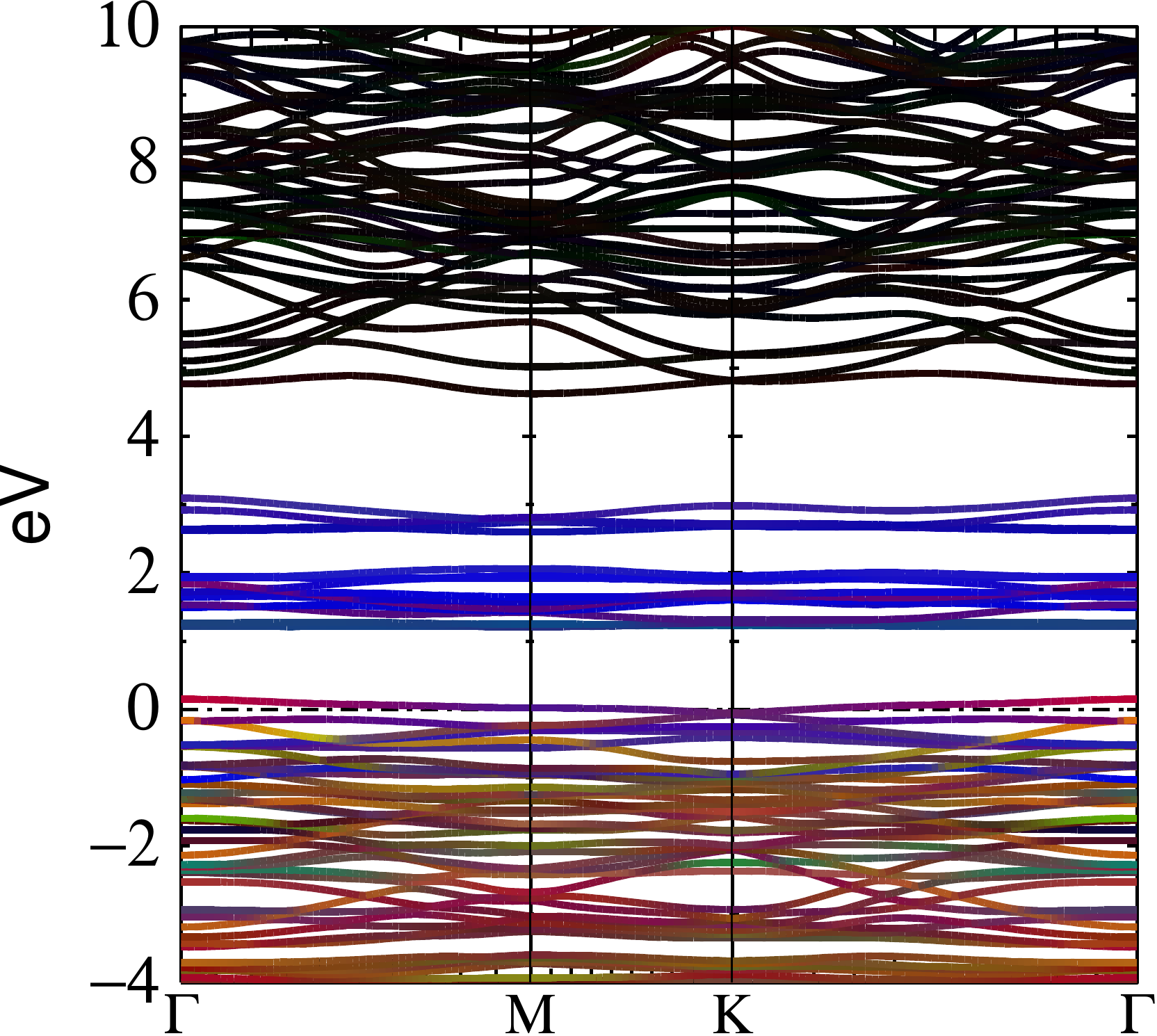}
	\includegraphics[width=0.32\textwidth, angle=-0]{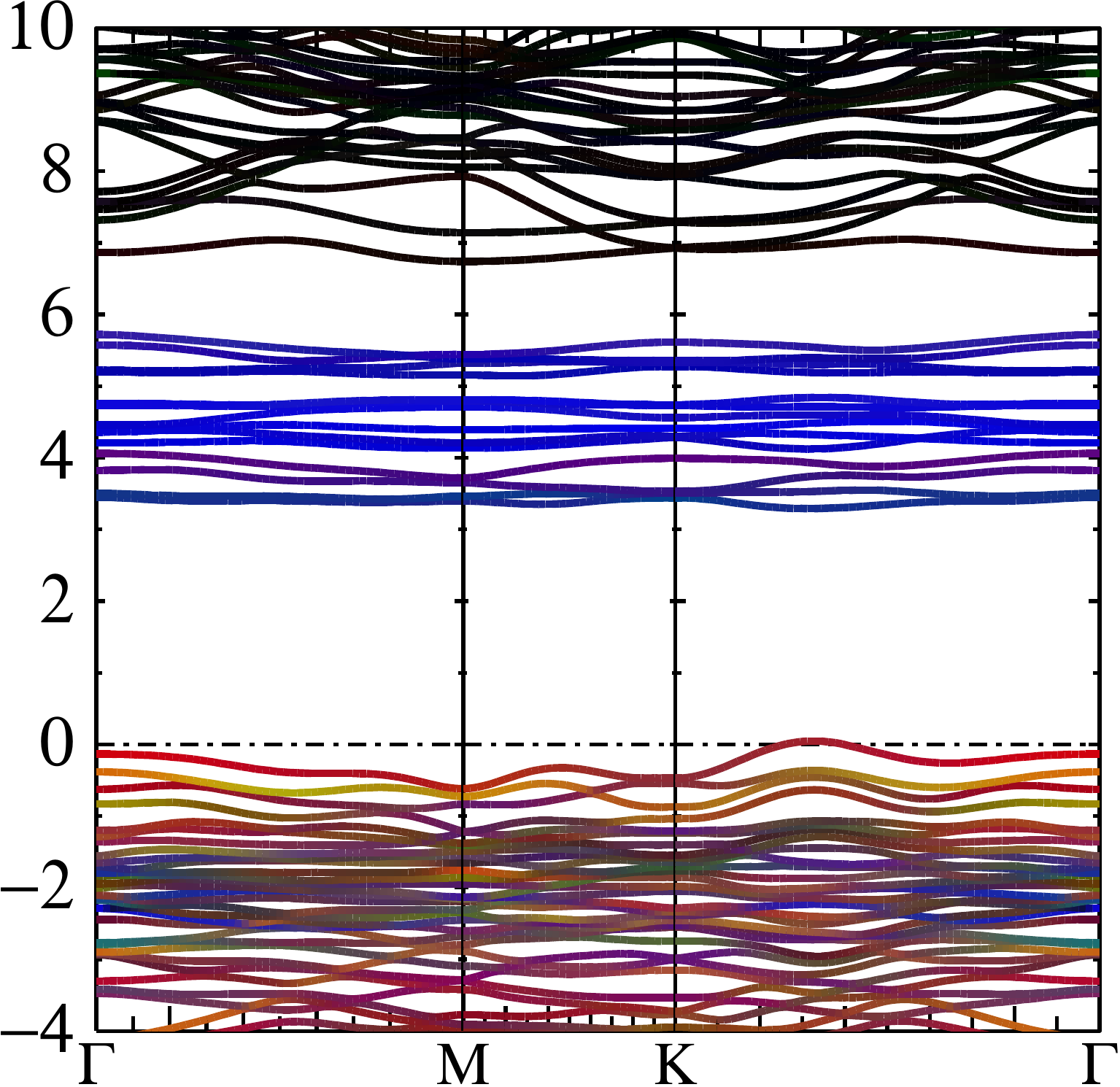}
	\includegraphics[width=0.32\textwidth, angle=-0]{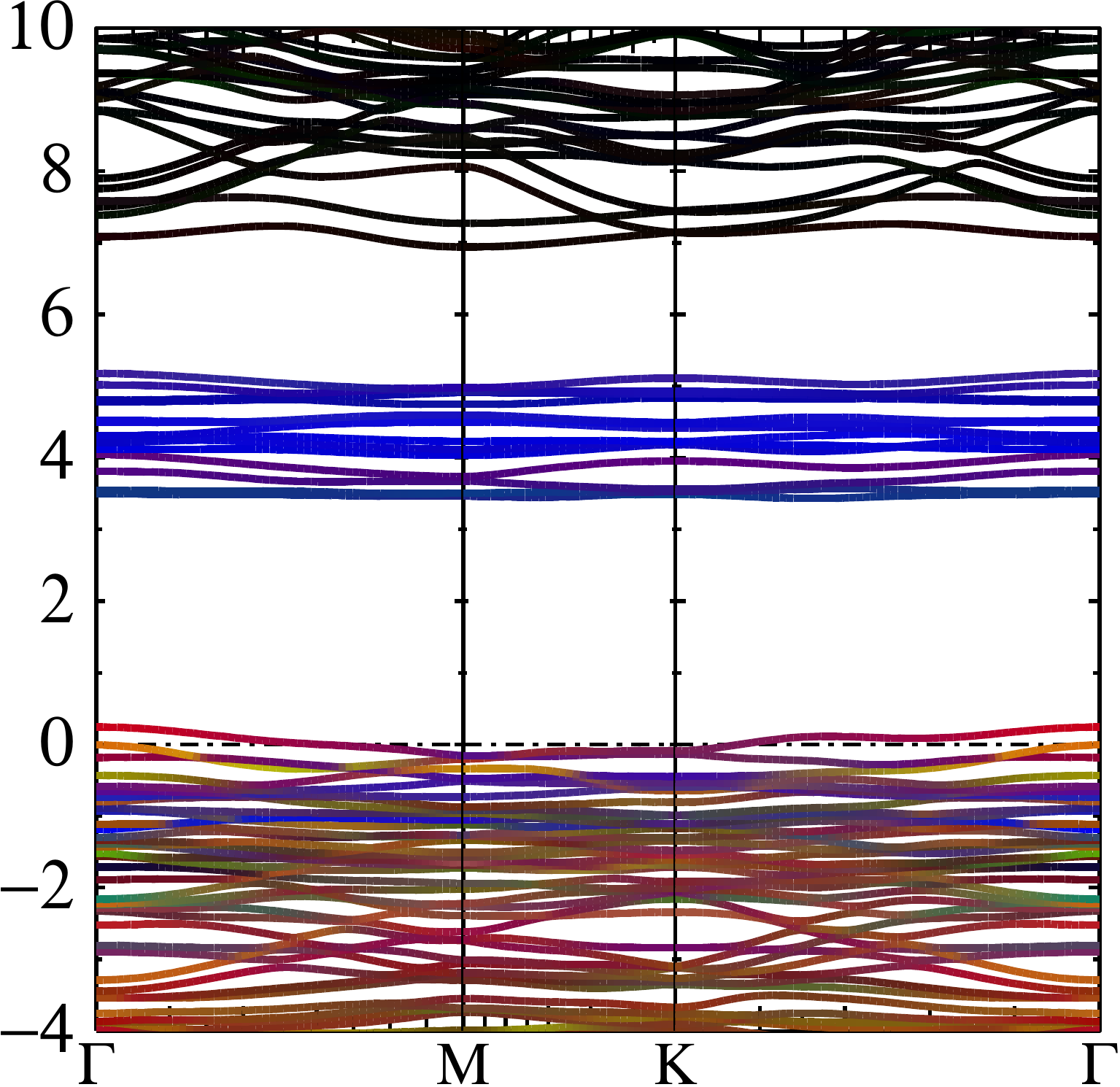}\\
	\caption{\ce{CrX3} : Top panels is for X= Cl, middle one for Br and bottom one for I. The colors correspond to X-$p_{x}+p_{y}$ (red), X p$_{z}$ (green), Cr-d (blue) (From left to right: DFT, QS\emph{GW} and QS$G\widehat{W}$ respectively).}
	\label{fig:BSFull}
			\end{center}
    \end{figure*}

    In Fig.~\ref{fig:BSFull} we show the band structures for all materials over larger energy windows. 

%

\end{document}